\documentclass[a4paper,12pt]{article}

\usepackage{amsfonts}
\usepackage{amsmath}
\usepackage[all]{xy}
\usepackage{verbatim}
\usepackage{rotating}

\usepackage[ps2pdf]{hyperref}

\hypersetup{pdftitle = {Permutation orbifolds  of $N=2$ supersymmetric minimal models}}
\hypersetup{pdfauthor = {Michele Maio, Bert Schellekens}}
\hypersetup{pdfsubject = {CFT-String Theory}}
\hypersetup{pdfkeywords = {Orbifolds, Extensions, Fixed Point Resolution, N=2 minimal models}}

\numberwithin{equation}{section}
\begin{document}
\title{Permutation orbifolds  of $N=2$ supersymmetric minimal models}
\author{M. Maio$^{1}$ and A.N. Schellekens$^{1,2,3}$\\~\\~\\
\\
$^1$Nikhef Theory Group, Amsterdam, The Netherlands
\\
~\\
$^2$IMAPP, Radboud Universiteit Nijmegen, The Netherlands
\\
~\\
$^3$Instituto de F\'\i sica Fundamental, CSIC, Madrid, Spain}

\def\Zbf{\bf Z}

\maketitle

\begin{abstract}
In this paper we apply the previously derived formalism of permutation orbifold conformal field theories to $N=2$ supersymmetric minimal models. By interchanging extensions and permutations of the factors we find a very interesting structure relating various conformal field theories that seems not to be known in literature. Moreover, unexpected exceptional simple currents arise in the extended permuted models, coming from off-diagonal fields. In a few situations they admit fixed points that must be resolved. We determine 
the complete CFT data with all fixed point resolution matrices for all simple currents of all $\mathbb{Z}_2$-permutations orbifolds of all
minimal $N=2$ models with $k\not=2\mod 4$.

\end{abstract}
\vskip -20cm
\hbox{ \hskip 12cm NIKHEF/2010-37  \hfil}


\clearpage
\tableofcontents

\section{Introduction}

Rational conformal field theory (RCFT) \cite{Belavin:1984vu} has proved to be a useful tool for
studying perturbative string theory, and especially string model
building. It provides a middle ground between approaches
based on free field theories (free bosons, free fermions and orbifolds)
on the one hand and geometric constructions on the other hand. 
While free field theory constructions are undoubtedly simpler and
more easily applicable to the computation of features of 
phenomenological interest such as couplings and  moduli dependence, 
there is a danger of them being too special. This may lead to incorrect
conclusions about what is possible or not in string theory, and
how generic certain features are. For example, the earliest attempts
to obtaining MSSM spectra using orbifold-based orientifolds were
plagued by chiral exotics \cite{Cvetic:2001tj}. 
However, the first detailed exploration of interacting RCFT orientifolds produced hundreds of thousands of distinct spectra without such unacceptable
particles \cite{Dijkstra:2004cc}. 

However, the set of RCFTs at our disposal is disappointingly small.
Decades ago it was conjectured that the moduli spaces of string theory
were densely populated by RCFT points, just as the $c=1$ moduli
space is densely populated by rational circle compactifications and
their orbifolds. 
Even today, it is not clear what
the status of that conjecture is, but it certainly does not have any
practical value. The only interacting RCFTs that we can really use for building
exact string are tensor products of $N=2$ minimal models, also known
as ``Gepner models" \cite{Gepner:1987qi,Gepner:1987vz}. Indeed, these are the only ones that have been used in orientifold
model building. In this situation we face the same risk mentioned above regarding
free CFTs: perhaps what we are finding is too special. For example, in a study of
the number of families in Gepner orientifolds it was found that the number three was
suppressed by a disturbing two to three orders of magnitude \cite{Dijkstra:2004cc} (similar conclusions were
obtained for $\mathbb{Z}_2 \times \mathbb{Z}_2$ orientifolds in \cite{Gmeiner:2005vz}). The origin of this
phenomenon is, to our knowledge, still not understood, but it would be interesting
to know if it persists beyond Gepner models.

The other area of application of RCFT model building, and the one where it was
historically used first, is the heterotic string. In this case a few results are
available beyond Gepner model building. This is possible because the computation
of the simplest heterotic spectra requires slightly less CFT data than what is
needed for orientifold spectra.

The full power of RCFT model building only manifests itself
if one uses the complete set \cite{Kreuzer:1993tf}
of simple current modular invariant partition functions (MIPFs)  \cite{Schellekens:1989am,Schellekens:1989dq} 
(See  \cite{Schellekens:1990xy} for a review of simple current MIPFs. The underlying symmetries were discovered independently in \cite{Intriligator:1989zw}).  Indeed, with only the trivial (diagonal or charge conjugation) MIPFs essentially nothing would have been found in orientifold model building. Indeed, already
basic physical constraints like world-sheet and space-time supersymmetry require a simple current MIPF.
Although the simple current symmetries can be read off from the modular transformation matrix $S$, and the
corresponding MIPFs can be readily constructed, often additional information is required
when the simple current action has fixed points \cite{Schellekens:1989uf,Schellekens:1990xy}.
In order to make full use of the complete simple current formalism we need the following data of the CFT
under consideration:
\begin{itemize}
\item{The exact conformal weights.}
\item{The exact ground state dimensions.}
\item{The modular transformation matrix $S$.}
\item{The fixed point resolution matrices $S^J$, for simple currents $J$ with fixed points.}
\end{itemize}
Not all of this information is needed in all cases. The matrices $S$ and $S^J$
are needed in the computation of boundary coefficients of orientifolds for simple current
MIPFs. In heterotic spectrum computations all we need to know is the first two items, plus the
simple current orbits implied by $S$. To compute the Hodge numbers of heterotic
compactifications, we only need to know the exact ground state dimensions of the Ramond
ground states.  

In addition to Gepner models, for which all this information is available, there are at least two other classes that are potentially usable: the Kazama-Suzuki models \cite{Kazama:1988qp}
and the
permutation orbifolds. In the former case, the coset construction gives us the
exact matrix $S$, and the results of \cite{Fuchs:1995tq} provide all the matrices $S^J$. The difficulty lies
in computing the exact CFT spectrum and the ground state dimensions. In some cases
this just requires the computation of branching functions, a tedious task that
can however be performed systematically. To our knowledge this has never been done, however.
In other cases, those with field identification
fixed points, no algorithm is currently available. In both cases, it has been possible
to compute at least the Hodge numbers for the diagonal MIPF (see respectively \cite{Font:1989qc} and \cite{Schellekens:1991sb}),
although in the latter case this
required some rather involved tricks to deal with fixed point resolution.

For permutation orbifolds the situation is more or less just the other way
around: it has been known for a long time how to compute their weights and ground state
dimensions, but there was no formalism for computing $S$ and $S^J$. Also in this case
it has been possible to compute the Hodge numbers and even the number of singlets for
the
diagonal invariants \cite{Klemm:1990df,Fuchs:1991vu}. However, meanwhile it as become
clear that the values of Hodge numbers offer a rather poor road map to the heterotic string landscape.
In particular they lead to the wrong impression that the number of families is large and very often a 
multiple of 4 or 6. The former problem disappears if one allows breaking of the gauge group 
$E_6$ to phenomenologically more attractive subgroups (ranging from $SO(10)$ via $SU(5)$
or Pati-Salam to just $SU(3)\times SU(2) \times U(1)$ (times other factors) by allowing asymmetric
simple current invariants \cite{Schellekens:1989wx,GatoRivera:2010gv}, whereas the second
problem can be solved by modifying the bosonic sector of the heterotic string, for example
by means of heterotic weight lifting \cite{GatoRivera:2010xn}, B-L lifting \cite{GatoRivera:2010xna} or $E_8$ breaking \cite{GatoRivera:2010xnb}.
All of these methods require knowledge of the full simple current structure of the building blocks.
This in its turn requires knowing $S$.

A first step towards the computation of $S$ for $\mathbb{Z}_2$ permutation orbifolds
was made in \cite{Borisov:1997nc}, almost ten years after permutation orbifolds were first  studied.
While this might seem sufficient for permutation orbifolds in heterotic string model building,
we will see that even in that case more is needed. For orientifold
model building with general simple current MIPFs certainly more is needed, as already mentioned above. The crucial ingredient
in both cases is fixed point resolution. Therefore we expect that significant progress can be made
by applying results we obtained recently \cite{Maio:2009kb,Maio:2009cy,Maio:2009tg}, 
extending the results of \cite{Borisov:1997nc} to fixed point resolution matrices $S^J$, for currents
$J$ of order 2. Since in $N=2$ minimal models all currents with fixed points have order 2, this seems
to be precisely what is needed. The purpose of this paper is to determine which of the CFT data
listed above
can now be computed for permutations orbifolds of $N=2$ minimal models, and provide algorithms for doing so.

\subsection{Basic concepts}

The easiest way of constructing rational conformal field theories is by taking the tensor product of existing ones. 
In the resulting CFT, all relevant CFT data is known from its factor theories. Another possibility is to take orbifolds. 
Orbifolds are already non-trivial theories, since they admit an untwisted and a twisted sector. 
The twisted sector is demanded by modular invariance. Normally the untwisted sector is easily derivable from the original theory, but twisted fields are much harder to determine.

There are many kind of orbifolds, depending on the particular model under consideration. In this paper we will consider the permutation orbifold, which arises when a tensor product CFT has at least two identical factor that can be permuted. 
The simplest instance of this orbifold is when there are only two identical factors to interchange. 
Start with the CFT $\mathcal{A}$ and build the tensor product $\mathcal{A}\otimes\mathcal{A}$. 
It has a manifest $\mathbb{Z}_2$ symmetry which flips the two factors. We denote this $\mathbb{Z}_2$ orbifold as 
\begin{equation}
\mathcal{A}\otimes\mathcal{A}/\mathbb{Z}_2\,.
\end{equation}
The spectrum was worked out for the first time in \cite{Klemm:1990df} using modular invariance, and twisted fields were determined. 
Subsequently, the modular $S$ and $T$ matrices were given in \cite{Borisov:1997nc} using an induction procedure on the algebra generators.

The next level of complication for a CFT is the simple-current extension. A simple current is a particular field of the theory, with simple fusion rules with any other field. 
If they have integer spin, they can be used to write down new modular invariant conformal field theories, known as 
simple current invariants. 
In the extension procedure, one computes the monodromy charge $Q_J$ of each field with respect to the 
simple current $J$ and organizes fields into orbits, keeping only those with integer monodromy. 
Algebraically, an extension is an orbifold projection, 
where one keeps states which are invariant under the monodromy operator $e^{2i\pi Q_J}$ and adds the twisted sector.

In principle the CFT data of such invariants are determined by those of the original theory, 
but the level of difficulty rapidly increases if there are fixed points, {\it i.e.} fields that the simple current leaves fixed under the fusion rules. 
Equivalently, fixed points are orbits with length one. If there are fixed points one needs a set of matrices ``$S^J$'' for each 
current $J$ acting on the fixed points \cite{Fuchs:1996dd}. Outside the fixed points of $J$,  $S^J$ vanishes. 
The full modular $S$ matrix is then expressed in terms of these $S^J$ matrices in a complicated way. 
Expressions for the $S^J$ matrix are known for WZW models, coset theories and 
extensions thereof \cite{Schellekens:1989uf,Fuchs:1996dd,Fuchs:1995zr,Schellekens:1999yg}.

When we combine extensions and permutation orbifolds, things become much more interesting and complicated at the same time. 
There it is always needed to worry about fixed point resolutions and $S^J$ matrices. 
The structure of simple currents and fixed points in the permutation orbifold was derived
in \cite{Maio:2009kb,Maio:2009cy} and also a unitary and modular invariant ansatz for the $S^J$ matrices exists \cite{Maio:2009tg}. 
Using the formula of \cite{Fuchs:1996dd}, we have checked that in simple current extensions these matrices $S^J$
yield a good $S$ matrix (satisfying the condition $(ST)^3=S^2$) and produce non-negative integer coefficients in the fusion rules. 

One may distinguish five kinds of fields in permutation orbifolds, which we will denote as follows. The labels $i$ and $j$ refer to primaries of the original CFT\footnote{We use a different notation for off-diagonal combinations than previous work 
\cite{Borisov:1997nc,Maio:2009kb,Maio:2009cy,Maio:2009tg}: $\langle i,j\rangle$ instead of $(i,j)$. This is to prevent confusion between the antisymmetric combination of the vacuum module, $(0,1)$, and the off-diagonal combination of fields nr. 0 and 1. The comma
 will be omitted in some cases. }:
 \begin{itemize}
\item Diagonal fields $(i,\xi)$, with $\xi=0$ or $1$. 
Here $\xi=0$ labels the symmetric combination and $\xi=1$ the anti-symmetric one. 
\item Off-diagonal fields $\langle i,j\rangle$, $i < j$.  
\item Twisted fields $\widehat{(i,\xi)}$, with $\xi=0$ or 1.
The $\widehat{(i,1)}$ denotes the excited twist field. 
\end{itemize}
Exact  formulas for the Virasoro characters of all these representations are known \cite{Borisov:1997nc}, and can be used
to get the exact ground state conformal weight and dimensions (see chapter 3 for further details).
  
Here we want to apply the results of \cite{Borisov:1997nc} and \cite{Maio:2009kb,Maio:2009cy,Maio:2009tg} to $N=2$ minimal models. This may seem to be
straightforward, as a supersymmetric CFT is just an example of a CFT, and the aforementioned results hold for {\it any} CFT. 
However,  the permutation orbifold obtained by applying \cite{Borisov:1997nc} turns out {\it not} to have world-sheet supersymmetry. This is related
to the fact that a straightforward Virasoro tensor product (the starting point for the permutation orbifold) does not have world-sheet supersymmetry
either, for the simple reason that tensoring produces combinations of R and NS fields. The solution to this problem in the case of the tensor product
is to extend the chiral algebra by a simple current of spin 3, the product of the world-sheet supercurrents of the two factors (or any two factors if there
are more than two). One might call this the supersymmetric tensor product.
But for this extended tensor product the formalism of \cite{Borisov:1997nc}  is not available. One can follow two paths to solve that problem: either one can 
try to generalize \cite{Borisov:1997nc} to supersymmetric tensor products (or more generally to extended tensor products) or one can try to
supersymmetrize the permutation orbifold. We will follow the second path.

One might expect that the chiral algebra of permutation orbifold has to be extended in order to restore world-sheet supersymmetry. That is indeed
correct, but it turns out that there are {\it two} plausible candidates for this extension: the symmetric and the anti-symmetric combination of the 
world-sheet supercurrent of the minimal model. Denoting the latter as $T_F$, the two candidates are the spin-3 currents $(T_F,0)$ and $(T_F,1)$.
Somewhat counter-intuitively, it is the second one that leads to a CFT with world-sheet supersymmetry. The first one,  $(T_F,0)$, gives rise
to a CFT that is similar, but does not have a spin-3/2 current of order 2. 

Both $(T_F,0)$ and $(T_F,1)$ have fixed points, but we know their resolution matrices from the general results of \cite{Maio:2009tg}. They
come in handy, because it turns out that one of these fixed points is the off-diagonal field $\langle 0,T_F \rangle$ of conformal weight $\frac32$. As stated above,
this is not a simple current of the permutation orbifold, but it is a well-known fact that chiral algebra extensions can turn primaries into
simple currents. This is indeed precisely what happens here. Since we know the fixed point resolution matrices of  $(T_F,0)$ and $(T_F,1)$ we
can work out the orbits of this new simple current. It turns out that in the former extension  $\langle 0,T_F \rangle$ has order 4, whereas in the latter it has
order 2. We conclude that the latter must be the supersymmetric permutation orbifold; we will refer to this CFT as ``$X$".
The fixed point resolution also  determines the action
of the new world-sheet supercurrent $\langle 0,T_F \rangle$ on all other fields, combining them into world-sheet superfields of either NS or R type. 

The current $\langle 0,T_F \rangle$ has no fixed points, as one would expect in an $N=2$ CFT (because it has two supercurrents of opposite charge, 
and acting with either one changes the charge). However, there are in general  more off-diagonal fields that turn into simple currents.
Some of these do have fixed points, and since the simple currents originate from fields that were not simple currents in the
permutation orbifold, our previous results do not allow us to resolve these fixed points. We find that this problem only occurs if $k=2 \mod 4$, where $k=1\ldots \infty$ is the integer parameter labelling the $N=2$ minimal models.

To prevent confusion we list here all the CFTs that play a r\^ole  in the story:

\begin{itemize}
\item{The $N=2$ minimal models.} 
\item{The tensor product of two identical $N=2$ minimal models. We will refer to this as $(N=2)^2$. } 
\item{The BHS-orbifold of the above. This is the permutation orbifold as described in \cite{Borisov:1997nc}. It will be denoted
$(N=2)^2_{\rm orb}$.} 
\item{The supersymmetric extension of the tensor product. This is the extension of the tensor product by the spin-3 current $(T_F,T_F)$. 
We will call this CFT $(N=2)^2_{\rm Susy}$ } 
\item{The supersymmetric permutation orbifold $(N=2)^2_{\rm Susy-orb}$. This is BHS orbifold extended by the spin-3 current $(T_F,1)$} 
\item{The non-supersymmetric permutation orbifold $X$. This is BHS orbifold extended by the spin-3 current $(T_F,0)$} 
\end{itemize}

\subsection{Plan}
The plan of this paper is as follows. \\
in section 2 we review the theory of $N=2$ minimal models, their spectrum and $S$ matrix. As far as the characters are concerned, we recall the coset construction and state a few known results from parafermionic theories, in particular the string functions.\\ In section 3 we
review general permutation orbifolds, the BHS formalism and its generalization to fixed point resolution matrices. 
Then in section 4 we move to the permutation orbifold of $N=2$ minimal models. We consider extensions by the various currents related to the spin-$\frac{3}{2}$ worldsheet supercurrent and explain how the exceptional off-diagonal currents appear.
We also work out the special extension of the orbifold by the symmetric and anti-symmetric representation of the worldsheet current. \\
In section 5 we study 
we study the exceptional simple currents and in particular the ones that have got fixed points. We give the structure of these off-diagonal currents as well as of their fixed points, in the case they have any. 
We illustrate the general ideas with the example of the minimal model at level two. In section 6 we summarize the orbit and fixed
point structures for the various CFTs we consider, and we also present the analogous results for $N=1$ minimal models, where
similar issues arise, and also some interesting differences. in section 7 we give our conclusions.

\section{$N=2$ minimal models}
In this section we review the minimal model of the $N=2$ super conformal algebra.

\subsection{The $N=2$ SCFT and minimal models}
The $N=2$ super conformal algebra (SCA) was first introduced in \cite{Ademollo:1976pp}. It contains the stress-energy tensor $T(z)$ (spin 2), a $U(1)$ current $j(z)$ (spin 1) and two fermionic currents $T_F^{\pm}(z)$ (spin $\frac{3}{2}$). In operator product form it reads:
\begin{subequations}
\label{SCA OPE}
\begin{eqnarray}
T(z) T(0) &\sim& \frac{c}{2z^4}+\frac{2}{z^2}T(0) +\frac{1}{z}\partial T(0) \\
T(z) T_F^{\pm}(0) &\sim& \frac{3}{2z^2}T_F^{\pm}(0) +\frac{1}{z}\partial T_F^{\pm}(0) \\
T(z) j(0) &\sim& \frac{1}{z^2}j(0) \frac{1}{z}\partial j(0) \\
T_F^{+}(z) T_F^{-}(0) &\sim& \frac{2c}{3z^3} +\frac{2}{z^2}j(0)+\frac{2}{z}T(0) +\frac{1}{z}\partial j(0) \\
T_F^{+}(z) T_F^{+}(0) &\sim& T_F^{-}(z) T_F^{-}(0)\sim 0\\
j(z) T_F^{\pm}(0) &\sim& \pm \frac{1}{z}T_F^{\pm}(0)\\
j(z) j(0) &\sim& \frac{c}{3z^2} \,.
\end{eqnarray}
\end{subequations}
Using the mode expansion
\begin{equation}
T(z)=\sum_{n\in\mathbb{Z}}\frac{L_n}{z^{n+2}}\,\,,\qquad
j(z)=\sum_{n\in\mathbb{Z}}\frac{J_n}{z^{n+1}}\,\,,\qquad
T_F^{\pm}(z)=\sum_{r\in\mathbb{Z}\pm\nu}\frac{G_{r}^{\pm}}{z^{r+\frac{3}{2}}}\,,
\end{equation}
the algebra (\ref{SCA OPE}) is equivalent to the (anti-)commutators
\begin{subequations}
\label{SCA modes}
\begin{eqnarray}
[L_m,L_n] &=& (m-n)L_{m+n}+\frac{c}{12}(m^3-m)\delta_{m,-n}\,\\
{[}L_m,J_n{]} &=&-n J_{m+n}\,, \\
\{G^+_r,G^-_s\} &=& 2L_{r+s}+(r-s)J_{r+s}+\frac{c}{3}(r^2-\frac{1}{4})\delta_{r,-s}\, \\
\{G^+_r,G^+_s\} &=& \{G^-_r,G^-_s\} =0\,, \\
{[}J_m,G^\pm_r{]} &=& \pm\frac{1}{c}G^\pm_{r+n}\,,\\
{[}J_m,J_n{]} &=& \frac{c}{3}m\delta_{m,-n} \,.
\end{eqnarray}
\end{subequations}
The shift $\nu$ can in principle be real, but for our considerations we take it to be integer (NS sector) or half-integer (R sector). 
Unitary representations of the $N=2$ SCA can exists for values of the central charge $c\geq3$ (infinite-dimensional representations) and for the discrete series $c<3$ (finite-dimensional representations). The latter ones are discrete conformal field theories, the $N=2$ minimal models, whose central charge is specified by an integer number $k$, called the level, according to:
\begin{equation}
\label{N=2 central charge}
c=\frac{3k}{k+2}\,.
\end{equation}
The Cartan subalgebra is generated by $L_0$ and $J_0$, hence primary fields are labelled by their weights $h$ and charges $q$:
\begin{equation}
L_0 |h,q \rangle =h |h,q\rangle\,\,,\qquad J_0 |h,q\rangle=q |h,q\rangle\,.
\end{equation}
The allowed values for $h$ and $q$ are given by 
\begin{equation}
\label{N=2 weights}
h_{l,m,s}=\frac{l(l+2)-m^2}{4(k+2)} +\frac{s^2}{8}\,\,,\qquad 
q_{m,s}=-\frac{m}{k+2}+\frac{s^2}{2}\,,
\end{equation}
where $l,\,m,\,s$ are integer numbers with the property that
\begin{itemize}
\item $l=0,\,1,\dots,\,k$
\item $m$ is defined ${\rm mod}\,\,2(k+2)$ (we will choose the range $-k-1\leq m\leq k+2$)
\item $s=-1,\,0,\,1,\,2$ mod $4$; $s=0,\,2$ for NS sector, $s=\pm1$ for R sector.
\end{itemize}
In addition, in order to avoid double-counting, one has to take into account that not all the fields are independent but are rather identified pairwise:
\begin{equation}
\phi_{l,m,s}\sim\phi_{k-l,m+k+2,s+2}\,.
\end{equation}

In order to be able to say something about the characters of the minimal model, let us mention the coset construction. The $N=2$ minimal models can be described in terms of the coset
\begin{equation}
\label{coset}
\frac{su(2)_k \times u(1)_2}{u(1)_{k+2}}\,.
\end{equation}
Throughout this paper, we use the convention that $u(1)_p$ contains $2p$ primary fields. 
The characters of this coset are decomposed according to
\begin{equation}
\chi^{su(2)_k}_l(\tau) \cdot \chi^{u(1)_2}_s (\tau)=
\sum_{m=-k-1}^{k+2}
\chi^{u(1)_{k+2}}_m(\tau) \cdot \chi_{l,m,s} (\tau)\,,
\end{equation}
where $\chi_{l,m,s}$ are the characters (branching functions) of the coset theory. 

\subsection{Parafermions}
We will soon see that $\chi_{l,m,s}$ will be determined in terms of the so-called \textit{string functions}, which are related to the characters of the \textit{parafermionic theories} \cite{Fateev:1985mm,Qiu:1986zf}. In order to determine $\chi_{l,m,s}$, let us consider $su(2)_k$ representations. Using the Weyl-Kac character formula \cite{Kac,Kac:1984mq}, $su(2)_k$ characters are given by a ratio of generalized theta functions:
\begin{equation}
\label{su2 characters}
\chi^{su(2)_k}_l(\tau,z)=
\frac{\Theta_{l+1,k+2}(\tau,z)+\Theta_{-l-1,k+2}(\tau,z)}{\Theta_{1,2}(\tau,z)+\Theta_{-1,2}(\tau,z)}\,,
\end{equation}
where by definition
\begin{equation}
\Theta_{l,k}(\tau,z)=\sum_{n\in\mathbb{Z}+\frac{l}{2k}}q^{k n^2}e^{-2 i \pi n k z}\,.
\end{equation}
Parafermionic conformal field theories are given by the coset
\begin{equation}
\frac{su(2)_k}{u(1)_k}\,, \qquad c=\frac{2(k-1)}{k+2}\,.
\end{equation}
We can decompose $su(2)_k$ characters in term of $u(1)_k$ and parafermionic characters as
\begin{equation}
\chi^{su(2)_k}_l(\tau,z) =
\sum_{m=-k+1}^{k}
\chi^{u(1)_k}_m(\tau,z) \cdot \chi^{{\rm para}_k}_{l,m} (\tau)\,.
\end{equation}
This decomposition also gives the weight of the parafermions:
\begin{equation}
h_{l,m}=\frac{l(l+2)}{4(k+2)}-\frac{m^2}{4k}\,\,,\qquad
l=0,1,\dots,k\,,\qquad m=-k+1,\dots,k\,.
\end{equation}
Using the fact that $u(1)_k$ characters are just theta functions,
\begin{equation}
\chi^{u(1)_k}_m(\tau,z)=
\frac{\Theta_{m,k}(\tau,z)}{\eta(\tau)}\,,
\end{equation}
the $su(2)_k$ characters become
\begin{equation}
\label{string function definition}
\chi^{su(2)_k}_m(\tau,z)=
\sum_{m=-k+1}^{k}
\frac{\Theta_{m,k}(\tau,z)}{\eta(\tau)}\cdot \chi^{{\rm para}_k}_{l,m} (\tau)
\equiv
\sum_{m=-k+1}^{k}
\Theta_{m,k}(\tau,z)\cdot C^{(k)}_{l,m} (\tau)\,,
\end{equation}
being $C^{(k)}_{l,m} (\tau)=\frac{1}{\eta(\tau)}\chi^{{\rm para}_k}_{l,m} (\tau)$ the $su(2)_k$ string functions. Here, $\eta(\tau)$ is the Dedekind eta function, which is a modular form of weight $\frac{1}{2}$,
\begin{equation}
\eta(\tau)=q^{\frac{1}{24}}\prod_{k=1}^{\infty}(1-q^k)\,,\qquad
\eta(\tau)^{-1}=q^{-\frac{1}{24}}\sum_{n=0}^{\infty}P(n)q^n\,,\qquad
q=e^{2i\pi\tau}\,,
\end{equation}
with $P(n)$ the number of partitions of $n$.

As an example, consider the case with $k=1$. Since the characters of $\chi^{su(2)_1}_m$ are the same as the characters of $\chi^{u(1)_1}_m$, we have
\begin{equation}
\chi^{{\rm para}_1}_{0,0}(\tau)=\chi^{{\rm para}_1}_{1,1}(\tau)=1\,\,,\qquad
\chi^{{\rm para}_1}_{0,1}(\tau)=\chi^{{\rm para}_1}_{1,0}(\tau)=0\,.
\end{equation}
These relations for $k=1$ generalize to arbitrary $k$ to give selection rules for the string functions. 
By decomposing $su(2)$ representations into $u(1)$ representations, the branching functions (i.e. the parafermions) should not carry $u(1)$ charge, since they correspond to the coset (\ref{coset}) where the $u(1)$ part has been modded out. Bearing this observation in mind, the general $su(2)_k$-character decomposition, including the selection rules, is
\begin{equation}
\chi^{su(2)_k}_l(\tau,z) =
\sum_{
{\small
\begin{array}{c}
 m=-k+1\\
l+m=0\,{\rm mod}\,2
\end{array}
}
}^k
C^{(k)}_{l,m} (\tau) \cdot \Theta_{m,k}(\tau,z) \,.
\end{equation}

\subsection{String functions and $N=2$ Characters}
The string functions of $su(2)_k$ are Hecke modular forms \cite{Kac:1984mq}. They can be expanded as a power sum with integer coefficients as
\begin{equation}
C^{(k)}_{l,m} (\tau)=\exp{\left[2i\pi\tau\left(\frac{l(l+2)}{4(k+2)}-\frac{m^2}{4k}-\frac{c}{24}\right)\right]}
\sum_{n=0}^{\infty}p_n q^n\,,
\end{equation}
with $c=\frac{3k}{k+2}$, where $p_n$ is the number of states in the irreducible representation with highest weight $l$ for which the value of $J_0^3$ and $N$ are $m$ and $n$. These integer coefficients depend in general on the string function labels $l$ and $m$ and are most conveniently extracted from the following expression\footnote{There exist many different ways of determining the $su(2)_k$ string functions. See for example \cite{Jacob:2000gw}, where a derivation is given in terms of representation theory of the parafermionic conformal models, or \cite{Jacob:2001rd}, where a new basis of states is provided for the parafermions. 
Our formula is the standard one, given in \cite{Kac:1984mq}. 
It also agrees with \cite{Nemeschansky:1989wg,Nemeschansky:1989rx}
For equivalent, but different-looking, expressions, see \cite{Fortin:2006dn,SchillingWarnaar}.
}:
\begin{equation}
\label{string function}
C^{(k)}_{l,m} (\tau)=\eta(\tau)^{-3} \sum_{-|x|<y\le |x|} {\rm sign}{(x)}\,e^{2i\pi\tau[(k+2)x^2-ky^2]}\,,
\end{equation}
where $x$ and $y$ belongs to the range
\begin{equation}
\label{range for string function}
(x,y)\,{\rm or}\,\left(\frac{1}{2}-x,\frac{1}{2}+y\right)\in \left(\frac{l+1}{2(k+2)},\frac{m}{2k}\right)+\mathbb{Z}^2\,.
\end{equation}
Equation (\ref{string function}) is actually \textit{the} solution to (\ref{string function definition}), when the l.h.s. is given as in (\ref{su2 characters}).

The string functions satisfy a number of properties, that can be proved by looking at (\ref{string function}) and at the summation range (\ref{range for string function}):
\begin{itemize}
\item $C^{(k)}_{l,m}=0$, if $l+m\neq 0$ mod $2$;
\item $C^{(k)}_{l,m}=C^{(k)}_{l,m+2k}$ , i.e. $m$ is defined mod $2k$;
\item $C^{(k)}_{l,m}=C^{(k)}_{l,-m}$;
\item $C^{(k)}_{l,m}=C^{(k)}_{k-l,k+m}$.
\end{itemize}

Using theta function manipulations, we can express the characters of the $N=2$ superconformal algebra in terms of the string functions as \cite{Gepner:1987qi,Blumenhagen:2009zz}
\begin{equation}
\label{MinChar}
\chi_{l,m,s}(\tau,z)=\sum_{j \,{\rm mod}\, k} C^{(k)}_{l,m+4j-s}(\tau)\cdot \Theta_{2m+(4j-s)(k+2),2k(k+2)}(\tau,kz)\,.
\end{equation}
This expression is invariant under any of the transformations $s\rightarrow s+4$ and $m\rightarrow m+2(k+2)$, which shows that $m$ is defined modulo $2(k+2)$ and $s$ modulo $4$. Also, $\chi_{l,m,s}=0$ if $l+m+s\neq 0$ mod 2 and moreover $\chi_{l,m,s}$ is invariant under the simultaneous interchange $l\rightarrow k-l$, $m\rightarrow m+k+2$ and $s\rightarrow s+2$. In the following, we will choose the standard range
\begin{equation}
l=0,\dots,k\,,\qquad m=-k-1,\dots,k+2\,,\qquad s=-1,\dots,2\
\end{equation}
for the labels of the $N=2$ characters. This range would actually produce an overcounting of states, since there is still the identification
\begin{equation}
\phi_{l,m,s}\sim \phi_{k-l,m+k+2,s+2}
\end{equation}
to take into account. For this purpose, it is more practical to consider the smaller range
\begin{itemize}
\item for k=odd:
\begin{equation}
\{0\le l<\frac{k}{2} \,,\,\forall m,\,\forall s\}
\end{equation}
\item for k=even:
\begin{eqnarray}
&&\{0\le l<\frac{k}{2} \,,\,\forall m,\,\forall s\}
\bigcup
\{l=\frac{k}{2} \,,\,m=1,\dots,k+1,\,\forall s\}\bigcup\\
&&\bigcup
\{l=\frac{k}{2} \,,\,m=0,\,s=0,1\}
\bigcup
\{l=\frac{k}{2} \,,\,m=k+2,\,s=0,1\}\nonumber
\end{eqnarray}
\end{itemize}
which automatically implements the above identification as well as the constraint $l+m+s=0$ mod $2$
\footnote{Observe however that formula (\ref{N=2 weights}) might give a negative weight for a field with labels $(l,m,s)$ in the range above. When this happens, we consider its identified primary with labels $(k-l,m+k+2,s+2)$, which is guaranteed to have positive weight.}. Taking this into account, the number of independent representations is given by
\begin{equation}
\#({\rm fields})=\underbrace{(k+1)}_{{\rm from}\,\,l}\cdot \underbrace{2(k+2)}_{{\rm from}\,\,m}
\cdot \underbrace{4}_{{\rm from}\,\,s}\cdot \underbrace{\frac{1}{2}}_{\rm ident.}
\cdot \underbrace{\frac{1}{2}}_{\rm constr.}=
2(k+1)(k+2)\,,
\end{equation}
while the number of simple currents is 
\begin{equation}
\#({\rm simple\,\,currents})=4\,(k+2)\,,
\end{equation}
in correspondence with all the fields having $l=0$ (as we will see in a moment).

To actually compute the minimal model characters using (\ref{MinChar}) is a complicated matter that can only
be done reliably using computer algebra. Results for the ground state dimensions are readily available in the
literature, but as we will see, this is not sufficient to determine the conformal weights and ground state dimensions of the permutation orbifolds. Since the number of characters of $N=2$ minimal models increases rapidly with $k$, it is not really practical to provide explicit character expansions in this paper. Therefore we will make them available electronically via the program {\tt kac} \cite{kac} that may also be used to compute all other CFT data discussed here.

\subsection{Modular transformations and fusion rules}
The coset construction has the additional advantage of making clear what the modular $S$ matrix is for the minimal models. It is just the product of the $S$ matrix of $su(2)$ at level $k$, the (inverse) $S$ matrix of $u(1)$ at level $k+2$ and  
the $S$ matrix of $u(1)$ at level $2$:
\begin{eqnarray}
\label{N=2 S matrix}
S^{N=2}_{(l,m,s)(l',m',s')}&=& 
S^{su(2)_k}_{l,l'} \left(S^{u(1)_{k+2}}_{m,m'}\right)^{-1} S^{u(1)_2}_{s,s'}=\\
&=&\frac{1}{2(k+2)} \sin{\left(\frac{\pi}{k+2}(l+1)(l'+1)\right)}\,
e^{-i\pi\left(\frac{ss'}{2}-\frac{mm'}{k+2}\right)}\,.\nonumber
\end{eqnarray}
The corresponding fusion rules \cite{Verlinde:1988sn} are
\begin{equation}
(l,m,s)\cdot(l',m',s')=\sum_{\lambda,\mu,\sigma}
N^{\lambda}_{\mu,\sigma}\,\delta^{(2(k+2))}_{m+m'-\mu,\,0}\,\delta^{(4)}_{s+s'-\sigma,\,0}\,\,
(\lambda,\mu,\sigma)\,,
\end{equation}
where $N^{\lambda}_{\mu,\sigma}$ are the $su(2)_k$ fusion coefficients. Here,  $\delta^{(p)}_{x,\,0}$ is equal to $0$, except if $x=0$ mod $p$ when it is $1$. As a consequence, all the fields $\phi_{0,m,s}$ (and only these) are simple currents, since they are all related to the identity of the $su(2)_k$ current algebra (or equivalently to the $su(2)_k$ representation with $l=k$, which is the only simple current of the $su(2)_k$ algebra). 
In particular, the field $T_F\equiv(0,0,2)$ (with $l=0$) will be relevant in the sequel. It has spin $\frac{3}{2}$ and multiplicity two: it contains the (two) fermionic generators $T^\pm(z)$ of the $N=2$ superconformal algebra.

\section{Permutation orbifold}
Before going into the details of the permutation orbifold of the $ N=2$ minimal models, let us recall a few properties of the generic permutation orbifold \cite{Borisov:1997nc}, restricted to the $\mathbb{Z}_2$ case
\begin{equation}
\mathcal{A}_{\rm perm} \equiv \mathcal{A}\times \mathcal{A}/\mathbb{Z}_2\,.
\end{equation}
If $c$ is the central charge of $\mathcal{A}$, then the central charge of $\mathcal{A}_{\rm perm}$ is $2c$.
The typical (for exceptions see below) weights of the fields are:
\begin{itemize}
\item $h_{(i,\xi)}=2 h_i$
\item $h_{\langle i,j\rangle}=h_i+h_j$
\item $h\widehat{(i,\xi)}=\frac{h_i}{2}+\frac{c}{16}+\frac{\xi}{2}$
\end{itemize}
for diagonal, off-diagonal and twisted representations. 
Sometimes it can happen that the naive ground state has dimension zero: then one must go to its first non-vanishing descendant whose weight is incremented by integers. 

For the sake of this paper, we are mostly interested in the orbifold characters. Let us recall the expressions of \cite{Borisov:1997nc} for the diagonal, off-diagonal and twisted $\mathbb{Z}_2$-orbifold characters. 
We denote by $\chi$ the characters of the original (mother) CFT $\mathcal{A}$ and by $X$ the characters of the permutation orbifold:
\begin{subequations}
\label{BHS characters}
\begin{eqnarray}
X_{\langle i,j \rangle}(\tau)&=&\chi_{i}(\tau)\cdot\chi_{j}(\tau) \\
X_{(i,\xi)}(\tau)&=&\frac{1}{2}\chi_{i}^2(\tau)+e^{i\pi\xi}\frac{1}{2}\chi_{i}(2\tau) \\
X_{\widehat{(i,\xi)}}(\tau)&=&\frac{1}{2}\chi_{i}(\frac{\tau}{2})+e^{-i\pi\xi}\,T_i^{-\frac{1}{2}}\,\frac{1}{2}\chi_{i}(\frac{\tau+1}{2})
\end{eqnarray}
\end{subequations}
where $T_i^{-\frac{1}{2}}=e^{-i\pi(h_i-\frac{c}{24})}$. \\
Now, each character in the mother theory can be expanded as
\begin{equation}
\chi(\tau)=q^{h_{\chi}-\frac{c}{24}}\,\sum_{n=0}^\infty d_n q^n \qquad \qquad ({\rm with}\,\,q=e^{2i\pi\tau})
\end{equation}
for some non-negative integers $d_n$. Observe that the $d_n$'s can be extracted from
\begin{equation}
\label{q-derivative}
d_n=\frac{1}{n!}\frac{\partial^n}{\partial q^n}\left.\left(\sum_{k=0}^\infty d_k q^k\right)\right|_{q=0}\,.
\end{equation}
Similarly, each character of the permutation orbifold can be expanded as
\begin{equation}
X(\tau)=q^{h_X-\frac{c}{12}}\,\sum_{n=0}^\infty D_n q^n
\end{equation}
for some non-negative integers $D_n$. A relation similar to (\ref{q-derivative}) holds for the $D_n$'s.

Using (\ref{BHS characters}) and (\ref{q-derivative}), we can immediately find the relationships between the $d_n$'s and the $D_n$'s. Here they are:
\begin{subequations}
\label{d-D relations}
\begin{eqnarray}
D^{\langle i,j \rangle}_k &=& \sum_{n=0}^k d_n^{(i)}\,d_{k-n}^{(j)} \\
D^{(i,\xi)}_k &=&  \frac{1}{2} \sum_{n=0}^k d_n^{(i)}\,d_{k-n}^{(i)}+
\left\{
\begin{array}{lr}
0 & {\rm if}\,\,k={\rm odd} \\
\frac{1}{2}\,e^{i\pi\xi}\,d_{\frac{k}{2}}^{(i)} & {\rm if}\,\,k={\rm even}
\end{array}
\right. \\
D^{\widehat{(i,\xi)}}_k &=& d_{2k+\xi}^{(i)}
\end{eqnarray}
\end{subequations}
These expressions are particularly interesting because they tell us that, if we want to have an expansion of the orbifold characters up to order $k$, then it is not enough to expand the original characters up to the same order $k$ (it would be enough for the untwisted fields), but rather we should go up to the higher order $2k+1$, as it is implied by the third line of (\ref{d-D relations}).

\def\ket#1{|#1\rangle}

There are two possible reasons why a ``naive" ground state dimension might vanish, so that the actual ground state weight is 
larger by some integer value. If a ground state $i$ has dimension one, the naive dimension of $(i,1)$ vanishes. Then the first non-trivial
excited state will occur for the  non-zero value of $d^{(i)}_n$.   Similarly, the conformal weight of an excited twist field $(\xi=1)$ is
larger than that of the unexcited one $(\xi=0)$ by half an integer, unless some odd excitations of the ground state vanish. In CFT, every
state $| \phi_i\rangle$, except the vacuum, always has an excited state $L_{-1} | \phi_i\rangle$. Furthermore, in $N=2$ CFTs even the
vacuum has an excited state $J_{-1} \ket{0}$. Therefore, in $N=2$ permutation orbifolds, the conformal weights of all ground states is
equal to the typical values given above, except when a state $\ket{i}$ has ground state dimension 1. Then the conformal weight
is larger by one unit.

Using these characters, one can compute their modular transformation and find the orbifold $S$ matrix. It was determined in 
\cite{Borisov:1997nc} and will be referred to as $S^{BHS}$. It reads as
\begin{subequations}
\label{BHS}
\begin{eqnarray}
S_{\langle mn \rangle(pq)}&=&S_{mp}\,S_{nq}+S_{mq}\,S_{np} \\
S_{\langle mn \rangle\widehat{(p,\chi)}}&=&0 \\
S_{\widehat{(p,\phi)}\widehat{(q,\chi)}}&=&\frac{1}{2}\,e^{2\pi i(\phi+\chi)/2} \,P_{ip} \\
S_{(i,\phi)(j,\chi)}&=&\frac{1}{2}\,S_{ij}\,S_{ij} \\
S_{(i,\phi)\langle mn \rangle}&=&S_{im}\,S_{in} \\
S_{(i,\phi)\widehat{(p,\chi)}}&=&\frac{1}{2}\,e^{2\pi i\phi/2} \,S_{ip} \,,
\end{eqnarray}
\end{subequations}
where the $P$ matrix (introduced in \cite{Bianchi:1990yu}) is defined by $P=\sqrt{T}ST^2S\sqrt{T}$.

For future reference, it is convenient to recall the ansatz for the $S^J$ matrices, as given in \cite{Maio:2009tg}:
\begin{subequations}
\label{MS ansatz}
\begin{eqnarray}
S^{(J,\psi)}_{\langle mn \rangle(pq)}&=&S^J_{mp}\,S^J_{nq}+(-1)^\psi S^J_{mq}\,S^J_{np} \\
S^{(J,\psi)}_{\langle mn \rangle\widehat{(p,\chi)}}&=&
\left\{
\begin{array}{cl}
0 & {\,\,\rm if\,\,}   J\cdot m=m\\
A\,S_{mp} & {\,\,\rm if\,\,}   J\cdot m=n
\end{array}
\right. \\
S^{(J,\psi)}_{\widehat{(p,\phi)}\widehat{(q,\chi)}}&=&
B\,\frac{1}{2}\,e^{i\pi\hat{Q}_J(p)}\,P_{Jp,q}\,e^{i\pi(\phi+\chi)} \\
S^{(J,\psi)}_{(i,\phi)(j,\chi)}&=&\frac{1}{2}\,S^J_{ij}\,S^J_{ij} \\
S^{(J,\psi)}_{(i,\phi)\langle mn \rangle}&=&S^J_{im}\,S^J_{in} \\
S^{(J,\psi)}_{(i,\phi)\widehat{(p,\chi)}}&=&C\,\frac{1}{2}\,e^{i\pi\phi} \,S_{ip}\,.
\end{eqnarray}
\end{subequations}
By modular invariance, the phases satisfy the following relations:
\begin{equation}
B=(-1)^\psi\,e^{3i\pi h_J}\,\,,\qquad A^2=C^2=(-1)^\psi\,e^{2 i\pi h_J}\,,
\end{equation}
$h_J$ being the weight of the simple current, which might depend on the central charge, rank and level of the original CFT. Note that $B$ is \textit{fully} fixed, while $A$ and $C$ are fixed \textit{up to a sign}. We choose the positive roots to get back the BHS $S$ matrix as special case.

\section{Permutations of $N=2$ minimal models}

In this section we consider the permutation orbifold of two $N=2$ minimal models at level $k$. The CFT resulting from modding out the $\mathbb{Z}_2$ symmetry in the tensor product $(N=2)_k\otimes(N=2)_k$ is known from \cite{Klemm:1990df,Fuchs:1991vu,Borisov:1997nc}. Here we focus mostly on the new interesting features arising when one extends the theory with various simple currents. 

As already mentioned, each $N=2$ minimal model at level $k$ admits a supersymmetric current $T_F(z)$ with 
ground state multiplicity equal to two and spin $h=\frac{3}{2}$. In the coset language, it corresponds to the NS field partner of the identity, namely $(l,m,s)=(0,0,2)$. This current transforms each NS field into its NS partner (with different $s$) and each R field into its R conjugate (corresponding to the other value of $s$). In order to see this, note that the $m$ and $s$ indices are just $u(1)$ labels, hence in the fusion of two representations they simply add up: $(s)\times(s')=(s+s'\,{\rm mod}\, 4)$ and $(m)\times(m')=(m+m'\,{\rm mod} \,2(k+2))$.

The field $T_F(z)$ has simple fusion rules with any other field and it generates two integer-spin simple currents in the permutation orbifold, corresponding to the symmetric and anti-symmetric representations $(T_F,0)$ and $(T_F,1)$ of diagonal-type fields, both with spin $h=3$. Both these currents can be used to extend the permutation orbifold. They are both of order two and, interestingly (but not completely surprisingly), their product gives back the anti-symmetric representation of the identity:
\begin{equation}
(T_F,0)\cdot (T_F,1)=(0,1)\,,
\end{equation}
with all the other possible products obtained from this one by using cyclicity of the order two. In other words, the fields $(0,0), (T_F,0), (0,1), (T_F,1)$ form a $\mathbb{Z}_4$ group under fusion.

We will study the extensions in the next two subsections, where we will also see the new CFT structure coming from interchanging extensions and orbifolds. Before we do this, however, let us first mention some generic properties of the orbifold. 
Consider the permutation orbifold of two $N=2$ minimal models at level $k$ and extend it by either the symmetric or the anti-symmetric representation of $T_F(z)$. The resulting theory has the old standard simple currents coming from $\phi_{0,m,s}$ (or equivalently $\phi_{k,m+k+2,s+2}$, by the identification) in the mother theory (in number equal to the number of simple currents of the $(N=2)_k$ minimal model and corresponding to the \textit{orbits} of their diagonal representations according to the fusion rules given in the next two subsections) and an equal number of \textit{exceptional} simple currents that were not simple currents before the extension (since coming from fixed off-diagonal orbits of $\phi_{0,m,s}$, as we will see below). 

The structure of the exceptional simple current is very generic: it is the same for both $(T_F,0)$ and $(T_F,1)$, so we can consider both here. The word exceptional means that they are simple currents just because their extended $S$ matrix satisfies the relation $S_{0J}=S_{00}$ \cite{Dijkgraaf:1988tf}. 
First of all, note that the orbifold simple currents come from symmetric and anti-symmetric representations of the mother simple currents, hence there are as many as twice the number of simple currents of the mother minimal theory. 
Secondly, all the exceptional currents correspond to the label $l=0$ (or equivalently $l=k$) as it should be, since related to the $su(2)_k$ algebra. This has the following consequence. Recall the orbifold (BHS) $S$ matrix in the untwisted sector \cite{Borisov:1997nc}:
\begin{eqnarray}
S^{BHS}_{(i,\psi)(j,\chi)}&=& \frac{1}{2}\,S_{ij}\,S_{ij}\nonumber\\
S^{BHS}_{(i,\psi)\langle m,n \rangle}&=& S_{im}\,S_{in}\nonumber
\end{eqnarray}
Using the minimal-model $S$ matrix (\ref{N=2 S matrix}) one has:
\begin{equation}
S_{(0,0,0)(0,0,0)}=
\frac{1}{2(k+2)}\,\sin{\left(\frac{\pi}{k+2}\right)}
=S_{(0,0,0)(0,m,s)}\nonumber
\end{equation}
and hence
\begin{equation}
\label{2S=S in BHS}
S^{BHS}_{((0,0,0),0),\langle(0,m,s),(0,m,s+2)\rangle}=2\,S^{BHS}_{((0,0,0),0),((0,0,0),0)}\,.
\end{equation}
This equality will soon be useful. In particular, the factor $2$ will disappear in the extension, promoting the off-diagonal fields $\langle(0,m,s),(0,m,s+2)\rangle$ into simple currents. We will come back later to these exceptional currents.

Let us show now that these exceptional simple currents of the $(T_F,\psi)$-extended orbifold correspond exactly to those particular \textit{off-diagonal fixed points} whose $(T_F,\psi)$-orbits ($\psi=0,\,1$) are generated from the simple currents of the mother $N=2$ minimal model.\\
Consider off-diagonal fields of the form $\langle(0,m,s),(0,m,s+2)\rangle$. They are fixed points of $(T_F,\psi)$, since\footnote{This is proved in the next subsections.} $T_F \cdot (0,m,s)=(0,m,s+2)$. The number of such orbits is equal to half the number of simple currents in the original minimal model (i.e. those fields with $l=0$). In the extension, they must be resolved. This means that each of them will give rise to two ``split'' fields in the extension. Hence their number gets doubled and one ends up with a number of split fields again equal to the number of simple currents of the original minimal model. Moreover, the extended $S$ matrix, $\tilde{S}$, will be expressed in terms of the $S^{J}$ matrix corresponding to $J\equiv (T_F,\psi)$, according to
\begin{equation}
\tilde{S}_{(a,\alpha)(b,\beta)}=
C\cdot[S^{BHS}_{ab}+(-1)^{\alpha+\beta}\,S^{(T_F,\psi)}_{ab}]\,.
\end{equation}
Recall that the $S^J$ matrix is non-zero only if the entries $a$ and $b$ are fixed points. The labels $\alpha$ and $\beta$ keep track of the two split fields ($\alpha,\,\beta=0\,,1$). The factor $C$ in front is a group theoretical quantity, that in case $a$ and $b$ are both fixed, is equal to $\frac{1}{2}$.

The generic formula for $S^J$ as given in \cite{Maio:2009tg} was recalled in (\ref{BHS}). In particular, the untwisted (i.e. diagonal and off-diagonal) entries of $S^J$ vanish, since $T_F$ does not have fixed points:
\begin{eqnarray}
S^{(T_F,\psi)}_{\langle m,n \rangle(p,q)} &=& S^{T_F}_{mp}\,S^{T_F}_{nq} +(-1)^\psi S^{T_F}_{mq}\,S^{T_F}_{np} \equiv 0\nonumber\\
S^{(T_F,\psi)}_{(i,\phi)(j,\chi)} &=& \frac{1}{2}\,S^{T_F}_{ij}\,S^{T_F}_{ij}\equiv 0\nonumber\\
S^{(T_F,\psi)}_{(i,\phi)\langle m,n \rangle} &=& S^{T_F}_{im}\,S^{T_F}_{in}\equiv 0\nonumber\,.
\end{eqnarray}
This implies that
\begin{equation}
\tilde{S}_{(a,\alpha)(b,\beta)}=
C\cdot S^{BHS}_{ab}
\end{equation}
for each split field corresponding to untwisted fixed points $a,\,b$. If either $a$ or $b$ are not fixed points, then $S^{(T_F,\psi)}$ is  automatically zero and the $\tilde{S}$ is given directly by $S^{BHS}$, up to the overall group theoretical factor $C$ in front, which is equal to $2$ if both $a$ and $b$ are not fixed points and $1$ if only one entry is fixed. Using (\ref{2S=S in BHS}), this implies that after fixed point resolution one would have
\begin{equation}
\tilde{S}_{((0,0,0),0)\langle(0,m,s),(0,m,s+2)\rangle_\alpha} = \tilde{S}_{((0,0,0),0)((0,0,0),0)}
\qquad (\alpha=0,\,1)\,.
\end{equation}
This means that
\begin{equation}
\label{exceptional simple currents}
\langle(0,m,s),(0,m,s+2)\rangle_\alpha\qquad \alpha=0,\,1
\end{equation}
are the exceptional simple currents in the extended theory, being $((0,0,0),0)$ the identity of the permutation orbifold and $(0,m,s)$ simple currents in the mother theory. The label $m$ runs over all the possible values, $m\in [-k-1,k+2]$; the label $s$ is fixed by the constraint $l+m+s=0$ mod $2$. This is the origin of the exceptional currents in the extended permutation orbifold of two $N=2$ minimal models. Note that, since in the off-diagonal currents both fields appear with $s$ and $s+2$, we can fix once and for all the $s$-labels in the exceptional currents to be $s=0$ in the NS sector and $s=-1$ in the R sector. 

These exceptional simple currents may have in principle fixed points. However, it turns out to be in general not the case: in fact, we will see that only four of the several exceptional currents have fixed points and only if $k=2$ mod $4$. We will come back to this later.

\subsection{Extension by $(T_F,1)$}
Since we will need it later, let us start by studying how the current under consideration, $(T_F,1)$, acts on different fields in the orbifold.
By looking at some specific examples or by computing the fusion rules, one can show that the orbits are given as in the following list. We denote the $N=2$ minimal representations as $i\equiv (l,m,s)$ and the ``shifted'' representations as $T_F\cdot i\equiv (l,m,s+2)$.
\begin{itemize}
\item Diagonal fields $(i,\xi)$ (recall that $\xi$ is defined mod $2$)
\begin{equation}
(T_F,1)\cdot (i,\xi) = (T_F\cdot i,\xi+1)
\end{equation}
\item Off-diagonal fields $\langle i,j \rangle$
\begin{equation}
(T_F,1)\cdot \langle i,j \rangle = \langle T_F\cdot i,T_F\cdot j \rangle
\end{equation}
\item Twisted fields $\widehat{(i,\xi)}$ (recall that $\xi$ is defined mod $2$)
\begin{eqnarray}
(T_F,1)\cdot \widehat{(i,\xi)} &=& \widehat{(i,\xi+1)}
\qquad {\rm if}\,\, i\,\,{\rm is}\,\, NS\,\,(s=0,\,2) \nonumber\\
&&\\
(T_F,1)\cdot \widehat{(i,\xi)} &=& \widehat{(i,\xi)}
\qquad {\rm if}\,\, i\,\,{\rm is}\,\, R\,\,(s=-1,\,1) \nonumber
\end{eqnarray}
\end{itemize}
A comment about possible fixed points is in order, since they get split in the extension and need to be resolved.
Observe that there cannot be any fixed points from the diagonal representations, 
since $T_F$ does not leave anything fixed. They will become all orbits and will all be kept in the extension, since they have integer monodromy:
\begin{equation}
Q_{(T_F,1)}(i,\xi)=2h_{T_F}+2h_i-2\left(h_i+\frac{1}{2}\right)\quad\in\mathbb{Z}\,.\nonumber
\end{equation}
The number of such orbits is equal to the number of fields in the mother minimal model.\\
On the other side, there are in general fixed points for off-diagonal and twisted representations.
The off-diagonal fixed points arise when $j=T_F\cdot i$, i.e. in our notation when $\langle i,j \rangle$ is of the
form $\langle(l,m,s),(l,m,s+2)\rangle$; the remaining off-diagonal fields organize themselves into orbits, of which some are kept and some are projected out, depending on their monodromy. In particular, using
\begin{equation}
Q_{(T_F,1)}\langle i,j\rangle=2h_{T_F}+(h_i+h_j)-(h_{T_F i}+h_{T_F j})\quad{\rm mod}\,\,\mathbb{Z}\,,\nonumber
\end{equation}
and the fact that, from the term $\frac{s^2}{8}$ in (\ref{N=2 weights}), $h_i-h_{T_F i}$ is $\frac{1}{2}$ if $i$ is NS and $0$ if $i$ is R, we see that the orbit $(\langle i,j\rangle, \langle T_F i,T_F j\rangle)$ is kept only if $i$ and $j$ are both NS or both R, otherwise they are projected out.\\
The twisted fixed points come from all the R representations and are kept in the extension,
while the twisted fields coming from NS representations are not fixed and projected out in the extension, since their monodromy charge 
\begin{equation}
Q_{(T_F,1)}\widehat{(i,\xi)}=2h_{T_F}+\widehat{(i,\xi)}-\widehat{(i,\xi+1)}\quad{\rm mod}\,\,\mathbb{Z}\nonumber
\end{equation}
is half-integer, being $(T_F,1)$ of integer spin and the difference of weights between a $\psi=0$-twisted field and a $\psi=1$-twisted field equal to $\frac{1}{2}$.

We will show soon that for $k=2$ mod $4$ some of the exceptional currents in the extension have fixed points. Let us say a few words about them. It turns out that these fixed points are either of the off-diagonal or twisted type: there are none of diagonal kind. To be slightly more concrete, they are specific $(T_F,1)$-orbits of off-diagonal fields plus all the twisted $(T_F,1)$-fixed points (necessarily corresponding to the Ramond fields of the original minimal model). We will not say more now, but will come back later. At the moment we are not able to resolve them: in other words, we do not know what their $S^J$ matrices are, $J$ denoting the particular exceptional currents.

One important exceptional currents of the permutation orbifold is the \textit{worldsheet supersymmetry} current, which is the only current of order two and spin $h=\frac{3}{2}$: it is the off-diagonal field coming from the tensor product of the identity with $T_F(z)$. It does not have fixed points, because $T_F$ does not. Let us denote it by $J^{w.s.}_{orb}\equiv \langle 0, T_F \rangle$. By the argument given above, $J^{w.s.}_{\rm orb}$ is guaranteed to be fixed by $(T_F,1)$. This means that in the extension it gets split into two fields, that we denote by $\langle 0, T_F \rangle_\alpha$, with $\alpha=0$ or $1$. In the appendix we check that indeed $\langle 0, T_F \rangle_\alpha$ has order two:
\begin{equation}
\langle 0, T_F \rangle_\alpha \cdot \langle 0, T_F \rangle_\alpha = (0,0)\,,
\end{equation}
where $(0,0)$ is the identity orbit.

Now consider the tensor product of two minimal models. We can either extend by $T_F(z)\otimes T_F(z)$ to make the product supersymmetric or we can mod out the $\mathbb{Z}_2$ symmetry and end up with the permutation orbifold. Let us start with the latter option. It is known \cite{Maio:2009tg} that one can go back to the tensor product by extending the orbifold by the anti-symmetric representation of the identity, $(0,1)$. What we do instead is extending the orbifold by $(T_F,1)$. The resulting theory is the $N=2$ supersymmetric permutation orbifold which has the worldsheet spin-$\frac{3}{2}$ current in its spectrum.

Alternatively, we can change the order and perform the extension before orbifolding. Note that each $N=2$ factor is supersymmetric, but the product is not. In order to make it supersymmetric, we have to extend it by the tensor-product current $T_F(z)\otimes T_F(z)$. As a result, in the tensor product only those fields survive whose two factors are either both in the NS or both in the R sector. In this way, the fields in the product have factors that are aligned to be in the same sector. Now we still have to take the $\mathbb{Z}_2$ orbifold. Starting from the supersymmetric product, by definition, we look for $\mathbb{Z}_2$-invariant states/combinations and add the proper twisted sector.
We will refer to this mechanism which transform the supersymmetric tensor product into the supersymmetric orbifold as \textit{super-BHS}, in analogy with the standard BHS from the tensor product to the orbifold. 
The following scheme summarizes this structure:
\begin{displaymath}
\xymatrix{
\boxed{(N=2)^2} \ar@/^/[d]^{BHS} \ar[r]^{T_F\otimes T_F} & \boxed{(N=2)^2_{\rm Susy}} \ar@/^/@2{->}[d]^{\rm super-BHS} \\
\boxed{(N=2)^2_{\rm orb}} \ar@/^/@{.>}[u]^{(0,1)} \ar[r]^{(T_F,1)\quad} & \boxed{(N=2)^2_{\rm Susy-orb}} \ar@/^/@{.>}[u]^{(0,1)}
}
\end{displaymath}

As a check, let us consider the following example. Take the case of level $k=1$. The $(N=2)_{1}$ minimal model has central charge equal to one and twelve primary fields (all simple currents). Its tensor product has central charge equal to two, as well as its $T_F\otimes T_F$-extension and $\mathbb{Z}_2$-orbifold. \\
By extending the tensor product by the current $T_F\otimes T_F$, one obtains the supersymmetric tensor product, with 36 fields. Instead, by going to the orbifold and extending by the current $(T_F,1)$, one obtains the supersymmetric orbifold with 60 fields. As a side remark, there is only one theory with this exact numbers of fields and same central charge and that is in addition supersymmetric, but only by working out the spectrum one can prove without any doubt that the theory in question is the $(N=2)_4$ minimal model, which is indeed supersymmetric.\\
We can continue now and extend the supersymmetric orbifold by the current $(0,1)$. This operation is the inverse of the $\mathbb{Z}_2$-orbifold (super-BHS). As expected, we end up to the supersymmetric tensor product. Equivalently, the $\mathbb{Z}_2$-orbifold of the supersymmetric tensor product gives back the supersymmetric orbifold, consistently.

\subsection{Extension by $(T_F,0)$}
Many things here are similar to the previous case. Let us start by giving the fusion rules of the current $(T_F,0)$ with any other field in the permutation orbifold.
\begin{itemize}
\item Diagonal fields $(i,\xi)$ (recall that $\xi$ is defined mod $2$)
\begin{equation}
(T_F,0)\cdot (i,\xi) = (T_F\cdot i,\xi)
\end{equation}
\item Off-diagonal fields $\langle i,j \rangle$
\begin{equation}
(T_F,0)\cdot \langle i,j \rangle = \langle T_F\cdot i,T_F\cdot j \rangle
\end{equation}
\item Twisted fields $\widehat{(i,\xi)}$ (recall that $\xi$ is defined mod $2$)
\begin{eqnarray}
(T_F,0)\cdot \widehat{(i,\xi)} &=& \widehat{(i,\xi)}
\qquad {\rm if}\,\, i\,\,{\rm is}\,\, NS\,\,(s=0,\,2) \nonumber\\
&&\\
(T_F,0)\cdot \widehat{(i,\xi)} &=& \widehat{(i,\xi+1)}
\qquad {\rm if}\,\, i\,\,{\rm is}\,\, R\,\,(s=-1,\,1) \nonumber
\end{eqnarray}
\end{itemize}
Again, the current $(T_F,0)$ does not have diagonal fixed points, but does have off-diagonal 
and twisted fixed points. The off-diagonal ones are like before, while the twisted ones come
this time from NS fields. Twisted fields coming from R representations are projected out in 
the extension. Each fixed point is split in two in the extended permutation orbifold and must be resolved. 
Moreover, there will also be orbits coming from the diagonal and off-diagonal fields.

Also for $(T_F,0)$-extensions a few exceptional currents might have fixed points.  They are either off-diagonal $(T_F,0)$-orbits or all the twisted $(T_F,0)$-fixed points (necessarily of Neveu-Schwarz origin).

As before, consider now the tensor product of two minimal models and its permutation orbifold. Extend the orbifold with the current $(T_F,0)$, i.e. the symmetric representation $T_F(z)$. One obtains a new, for the moment mysterious, CFT that we denote by $X$. $X$ is not supersymmetric, since it does not contain the worldsheet supercurrent of spin $h=\frac{3}{2}$. To be more precise, $X$ does contain a spin $\frac{3}{2}$-current, which is again the off-diagonal field $\langle 0, T_F \rangle$. However, it is \textit{not} the worldsheet supersymmetry current. The reason is that in this case $\langle 0, T_F \rangle$ (or rather the two split fields $\langle 0, T_F \rangle_\alpha$, with $\alpha=0$ or $1$) has order $4$, instead of order $2$: acting twice with $J^{w.s.}_{\rm orb}(z)$ we should get back to the same field, but we do not. As we prove in the appendix:
\begin{equation}
\langle 0, T_F \rangle_\alpha \cdot \langle 0, T_F \rangle_\alpha = (0,1)\,,
\end{equation}
with $(0,1)\cdot(0,1)=(0,0)$. Hence there is no such a current as $J^{w.s.}_{\rm orb}(z)$ in $X$. Continuing extending this time by the current $(0,1)$ we get back to the familiar theory $(N=2)^2_{\rm Susy}$. 
The summarizing graph is below:
\begin{displaymath}
\xymatrix{
\boxed{(N=2)^2} \ar@/^/[d]^{BHS} \ar[r]^{T_F\otimes T_F} & \boxed{(N=2)^2_{\rm Susy}}  \\
\boxed{(N=2)^2_{\rm orb}}  \ar@/^/@{.>}[u]^{(0,1)} \ar[r]^{(T_F,0)\quad} & \boxed{{\rm Non-Susy}\,\,X} \ar[u]^{(0,1)}
}
\end{displaymath}

\subsection{Common properties}
By looking at the two graphs, we notice that there are two distinct ways of reproducing the behavior of the current $T_F\otimes T_F$ which makes the tensor product of two minimal models supersymmetric. We can go either through the supersymmetric permutation orbifold or through the more mysterious non-supersymmetric CFT $X$, as is shown below.
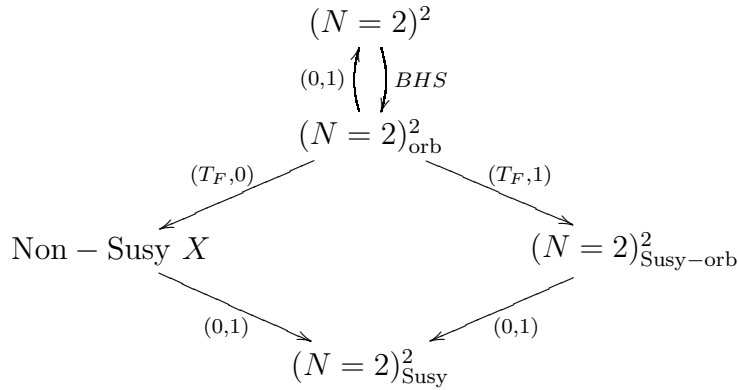
\begin{displaymath}
\xymatrix{
& (N=2)^2 \ar@/^/[d]^{BHS} & \\
& \ar@/^/[u]^{(0,1)} \ar[dl]_{(T_F,0)} (N=2)^2_{\rm orb} \ar[dr]^{(T_F,1)} &\\
{\rm Non-Susy}\,\,X \ar[dr]_{(0,1)} & & (N=2)^2_{\rm Susy-orb} \ar[dl]^{(0,1)} \\
& (N=2)^2_{\rm Susy} & \\
}
\end{displaymath}
We can summarize the commutativity of this diagram as:
\begin{equation}
(T_F\otimes T_F) \circ (0,1) = (0,1) \circ (T_F,\psi)
\end{equation}
when acting on $(N=2)^2_{\rm orb}$. The small circle $\circ$ means composition of extensions, e.g. $(J_2 \circ J_1) \mathcal{A}$ means that we start with the CFT $\mathcal{A}$, then we extend it by the simple current $J_1$ and finally we extend it again by the simple current $J_2$.

It is useful to ask what happens to the exceptional current $\langle 0, T_F \rangle$ (which coincides with $J^{w.s.}_{\rm orb}(z)$ for the $(T_F,1)$-extension). Using the fusion rules given earlier, it is easy to see that $\langle 0, T_F \rangle$ is fixed by both $(T_F,0)$ and $(T_F,1)$, because of the shift by $T_F$ in both the factors in off-diagonal fields and the symmetrization of the tensor product. As a consequence, the fixed point resolution is needed in both situations for the field $\langle 0, T_F \rangle$. 

Let us make a comment on the nature of the CFT $X$. We have already stressed enough that it is not supersymmetric. However, by looking at it more closely, it is quite similar to the supersymmetric orbifold $(N=2)^2_{\rm Susy-orb}$. For example, they contain the same number of fields and in particular they have the same diagonal and off-diagonal fields. They only differ for their twisted fields, being of R type in the supersymmetric orbifold and of NS type in $X$.

Another interesting point is that the $(0,1)$ extension of both $X$ and $(N=2)^2_{\rm Susy-orb}$ gives back the same answer, namely the $(N=2)^2_{\rm Susy}$. One could ask how this happens in detail. The reason is that, after the $(T_F,\psi)$-extension (either $\psi=0$ or $1$) of the orbifold, one is left with orbits and/or fixed points corresponding to orbifold fields of diagonal, off-diagonal and twisted type. In particular, as we already mentioned before, from the twisted fields only the fixed points survive, with the difference that for $\psi=1$ they come from the Ramond sector and for $\psi=0$ from the NS sector. However, they are completely projected out by the $(0,1)$-extension, which leaves only untwisted (i.e. off-diagonal and diagonal -both symmetric and anti-symmetric-) fields in the supersymmetric tensor product\footnote{The reason is that the current $(0,1)$ always couples a twisted field $\widehat{(p,0)}$ to its partner $\widehat{(p,1)}$, as it is shown in the appendix. Since these fields have weights which differ by $\frac{1}{2}$, then their monodromy will be half-integer and they will be projected out in the $(0,1)$-extension.}.

\section{Exceptional simple currents and their fixed points}
Let us be a bit more precise on the exceptional simple currents which admit fixed points. There are four of them and they are always related to the following mother-theory simple currents 
\begin{equation}
J_+\equiv(l,m,s)\equiv(0,\frac{k+2}{2},s)\equiv(k,-\frac{k+2}{2},s+2)
\end{equation}
and
\begin{equation}
J_-\equiv (0,-\frac{k+2}{2},s)\equiv(k,\frac{k+2}{2},s+2)
\end{equation}
(with $s=0$ in the NS sector, $s=-1$ in the R sector). We will soon prove that $s$ must be in the NS sector. i.e. $s=0$, otherwise there are no fixed points. Using the facts that $m$ is defined mod $2(k+2)$ and that $s$ is defined mod $4$, together with the identification $(l,m,s)=(k-l,m+k+2,s+2)$, it is easy to show that $J_+$ and $J_-$ are of order four, i.e. $J_+^4=J_-^4=1$. 
Moreover, we will soon show that off-diagonal fixed points of the exceptional currents originate from fields in the mother $N=2$ theory with $l$-label equal to $l=\frac{k}{2}$. One can easily check that, on these fields, the square of $J_\pm$, $J^2_\pm$, acts as follows. 
For $J_\pm$ in the R sector, $J_\pm^2$ fixes any other field (either R or NS) of the original minimal model:
\begin{equation}
(J_\pm\in R)\quad J_\pm^2:\,\, (l=\frac{k}{2},m,s) \longrightarrow (l=\frac{k}{2},m,s)
\Longrightarrow J_\pm^2\simeq 1\equiv (0,0,0)\,,
\end{equation}
acting on them effectively as the identity; for $J_\pm$ in the NS sector, $J_\pm^2$ takes an R (NS) field into its conjugate R (NS) field:
\begin{equation}
(J_\pm\in NS)\quad J_\pm^2:\,\, (l=\frac{k}{2},m,s) \longrightarrow (l=\frac{k}{2},m,s+2)
\Longrightarrow J_\pm^2\simeq T_F\equiv (0,0,2)\,,
\end{equation}
acting effectively as the supersymmetry current.

Having introduced now the currents $J_\pm$ in the mother theory, we can write down the four simple currents in the orbifold theory extended by $(T_F,\psi)$ which admit fixed points. Recalling that $T_F=(0,0,2)$ acts by shifting by two the $s$-labels in the original minimal model, we can consider the following off-diagonal fields in the permutation orbifold:
\begin{equation}
\label{formual for simple currents in orb}
\langle J_\pm,T_F\cdot J_\pm \rangle\,.
\end{equation}
The two off-diagonal combinations above satisfy the condition (\ref{2S=S in BHS}); hence, after fixed point resolution, each of them generates two exceptional simple currents (for a total of four) in the $(T_F,\psi)$-extended theory:
\begin{equation}
\label{formual for exceptional simple currents}
\langle J_\pm,T_F\cdot J_\pm \rangle_\alpha\,\,,\qquad \alpha=0,\,1\,,
\end{equation}
being $T_F\cdot J_\pm =(0,\pm \frac{k+2}{2},s+2)$. This is another way of re-writing (\ref{exceptional simple currents}), specialized to the exceptional currents that have fixed points.

If one wants to be very precise about the fixed points, one should study the fusion coefficients, which is in the present case very complicated, but in principle doable. However, we can still make some preliminary progress using intuitive arguments. First of all, since the resolved currents (\ref{exceptional simple currents}) carry an index $\alpha$ which distinguishes them, but are very similar otherwise, it is reasonable to expect that they might have the same fixed points and that hence the fixed-point conformal field theories corresponding to the exceptional currents might be pairwise identical. This is indeed what happens.\\
Secondly, observe that in (\ref{exceptional simple currents}) the field $(0,m,s)$ is equivalent to $(k,m+k+2 \,\,{\rm mod}\,\,2(k+2),s+2\,\, {\rm mod}\,\, 4)$. From the $su(2)_k$ algebra, the field labelled by $l=k$ is the only non-trivial simple current with fusion rules given by
\begin{equation}
(k)\cdot(j)=(k-j)\,,
\end{equation}
so in order for it to have fixed points, $k$ must be at least even. Moreover, $j$ is a fixed point of the $su(2)_k$ algebra if and only if $j=\frac{k}{2}$. This argument tells us that off-diagonal fixed points of (\ref{exceptional simple currents}) must be orbits whose component fields have $l$-label equal to $l=\frac{k}{2}$. This is indeed what happens.

Actually there are only four (coming from the above two resolved) exceptional simple currents which have fixed points and the corresponding four fixed-point conformal field theories are pairwise identical. Indeed, the exceptional simple currents have $m$-label equal to $m=\pm\frac{k+2}{2}$, even $s$-label and hence the generic constraints $l+m+s=0 \,\,{\rm mod} \,\,2$ implies that $k=2$ mod $4$. 

Let us describe more in detail the exceptional simple currents with fixed points. Consider again (\ref{formual for exceptional simple currents}) and study the fusion rules of (\ref{formual for simple currents in orb}). We are most interested in off-diagonal fixed points, because they have an interesting structure; as far as the other kind (namely twisted) of fixed points is concerned, they are as already reported in the previous section (namely of NS type for $(T_F,0)$ and of R type for $(T_F,1)$). Compute the fusion rule of the current $(J_\pm,T_F J_\pm)$ with any field of the form:
\begin{equation}
\label{offdiag f.p. of except curr}
\langle f,J_\pm f'\rangle\,,
\end{equation}
where $f'$ has either the same $s$-label as $f$ or different; in other words, either $f'=f$ or $f'=T_F f$. Here, $f$ and $f'$ label primaries of the original $N=2$ minimal model which might be fixed points of (\ref{formual for exceptional simple currents}), having their $l$-values equal to $l=\frac{k}{2}$. Explicitly, $f=(\frac{k}{2},m,s)$ and $f'=(\frac{k}{2},m,s')$,with $s'=s$ or $s'=s+2$.

We would like to show that the fields $\langle f,J_\pm f'\rangle$ constitute the subset of off-diagonal fixed points for the exceptional currents. For most of them, this subset will be empty, but not for (\ref{formual for exceptional simple currents}). As a remark, note that not all the fields in (\ref{offdiag f.p. of except curr}) are independent, since they are identified pairwise by the extension. We will come back to this at the end of this subsection.

Now let us compute the fusion rules. Naively:
\begin{eqnarray}
\langle J_\pm,T_F J_\pm\rangle \cdot \langle f,J_\pm f'\rangle&\propto&
(J_\pm\otimes T_F J_\pm+T_F J_\pm\otimes J_\pm)\cdot
(f\otimes J_\pm f'+J_\pm f'\otimes f)\nonumber\\
&=&
(J_\pm f\otimes T_F f'+J_\pm^2 f'\otimes T_F J_\pm f +\nonumber\\
&&
\qquad\qquad\qquad +T_F J_\pm f\otimes J_\pm^2 f'+T_F J_\pm^2 f'\otimes J_\pm f)\,.\nonumber
\end{eqnarray}
For currents in the R sector, $J_\pm^2=1$, while $J_\pm^2=T_F$ in the NS sector; hence the above expression simplifies in both cases:
\begin{equation}
\langle J_\pm,T_F J_\pm\rangle\cdot\langle f,J_\pm f'\rangle \propto \dots=
\left\{
\begin{array}{cc}
(J_\pm f\otimes T_F f'+ f'\otimes T_F J_\pm f +& {\rm R\,\,sector}\\
\qquad+T_F J_\pm f\otimes f'+T_F f'\otimes J_\pm f)\,.\nonumber & \\
&\\
(J_\pm f\otimes f'+ T_F f'\otimes T_F J_\pm f +& {\rm NS\,\,sector}\\
\qquad+T_F J_\pm f\otimes T_F f'+f'\otimes J_\pm f) &
\end{array}
\right.
\nonumber
\end{equation}
In terms of representation, we can decompose the r.h.s. in two pieces corresponding to the following symmetric representations:
\begin{eqnarray}
\label{temp fp of excep curr}
({\rm R})\quad \langle J_\pm,T_F J_\pm\rangle \cdot \langle f,J_\pm f'\rangle&=& 
\langle J_\pm f,T_F f'\rangle+\langle f',T_F J_\pm f\rangle\nonumber\\
({\rm NS})\quad \langle J_\pm,T_F J_\pm\rangle \cdot\langle f,J_\pm f'\rangle&=& 
\langle f',J_\pm f\rangle+\langle T_F f',T_F J_\pm f\rangle 
\end{eqnarray}
We have replaced here the proportionality symbol with an equality: a more accurate calculation of the fusion coefficients would show that the proportionality constant is indeed one. 
It is crucial that none of the two pieces in the first line (R sector) reduces to $(f,J_\pm f')$ as on the l.h.s.; on the contrary, either of them does, respectively if $f=f'$ and $f'=T_F f$, in the second line (NS sector). For example, in the NS situation, this is obvious in the case $f=f'$; if $f'=T_F f$ instead, we  must remember that the  brackets means symmetrization and that off-diagonal fields that are equal up to the action of $(T_F,\psi)$ are actually \textit{identified} by the extension. Similar arguments hold for the R situation as well.

Note here that the two pieces in (\ref{temp fp of excep curr}) are related by the application of $T_F$: if we talked about tensor product fields then the relation would be given by the tensor product $T_F\otimes T_F$, but since we are working in the orbifold, it is actually provided by the diagonal representation $(T_F,\psi)$. Let us move now to the extended orbifold. 

From the fusion rules given earlier, in the permutation orbifold extended by $(T_F,\psi)$, off-diagonal fields belong to the same orbit if and only if
\begin{equation}
(T_F,\psi)\cdot\langle i,j \rangle=\langle T_F i,T_F j\rangle\,.
\end{equation}
Since
\begin{equation}
(T_F,\psi)\cdot \langle f,J_\pm f'\rangle=\langle T_F f,T_F J_\pm f'\rangle\,,
\end{equation}
then the two quantities appearing on the r.h.s. of (\ref{temp fp of excep curr}) are identified by the extension and add up to give
\begin{eqnarray}
\label{fp of excep curr}
({\rm R})\qquad \langle J_\pm,T_F J_\pm\rangle\cdot\langle f,J_\pm f'\rangle&=& \langle J_\pm f,T_F f'\rangle\,,\nonumber\\
({\rm NS})\qquad \langle J_\pm,T_F J_\pm\rangle\cdot\langle f,J_\pm f'\rangle &=& \langle f',J_\pm f\rangle\,.
\end{eqnarray}
As a consequence, exceptional currents coming from R fields never have fixed points (neither if $f=f'$ nor if $f'=T_F f$), while NS fields do have. 
This shows that the exceptional simple currents with fixed points arise only for NS fields in the mother theory and they are exactly of the desired form.

As a consistency check, let us give the following argument about the currents (\ref{formual for exceptional simple currents}) (equivalently, identify $l\rightarrow k-l\,,\dots$ etc). We have already established that $k$ must be even in order for the currents to have fixed points, so we can discuss the two options of $k=4p$ and $k=2+4p$ (for $p\in\mathbb{Z}$) separately. In the former case, $k=4p$,
\begin{equation}
h_{\langle J_\pm,T_F\cdot J_\pm\rangle_\alpha}=h_{J_\pm}+h_{T_F\cdot J_\pm}=2\cdot\frac{3k}{16}=\frac{3p}{2}\,.
\end{equation}
This is either integer or half-integer, depending on $p$, so the currents might admit fixed points. However, the current $m$-label is equal to $2p+1\in\mathbb{Z}_{\rm odd}$; since the $l$-label is even, then the $N=2$ constraint forces the $s$-label to be $\pm 1$. As a consequence, the currents (\ref{formual for exceptional simple currents}) are of Ramond-type and hence cannot have fixed points.
In the latter case, $k=2+4p$,
\begin{equation}
h_{\langle J_\pm,T_F\cdot J_\pm\rangle_\alpha}=h_{J_\pm}+h_{T_F\cdot J_\pm}=
\left(\frac{3k}{16}-\frac{1}{8}\right)+\left(\frac{3k}{16}+\frac{3}{8}\right)=1+\frac{3p}{2}\,.
\end{equation}
This is either integer or half-integer, depending on $p$, then the current can have fixed points. Moreover, since the $m$-label is equal to $2p+2\in\mathbb{Z}_{\rm even}$, the currents (\ref{formual for exceptional simple currents}) are now of NS-type, hence they \textit{will have} fixed points.

Needless to say, we do expect all a priori possible fields of the form (\ref{offdiag f.p. of except curr}) to survive the $(T_F,\psi)$-extension, the reason being that their $(T_F,\psi)$-orbits must have zero monodromy charge with respect to the current $(T_F,\psi)$. As an exercise, let us compute this charge and prove that it vanishes (mod integer). For this purpose, we need to know the weight of (\ref{offdiag f.p. of except curr}). Since
\begin{equation}
h_{J_\pm f}=h_f-\frac{1}{16}(k+2\pm 4m)
\end{equation}
$m$ being the $m$-label of the field $f$, then 
\begin{equation}
h_{\langle f,J_\pm f'\rangle}=h_f+h_{J_\pm f'}=2h_f-\frac{1}{8}(k+2\pm 4m)+\frac{1}{2}\,\delta_{f',T_F f}\,.
\end{equation}
Similarly, we need to compute $h_{T_F f,T_F J_\pm f'}$. Since 
\begin{equation}
h_{T_F J_\pm f}=h_{T_F f}-\frac{1}{16}(k+2\pm 4m)
\end{equation}
then again 
\begin{equation}
h_{\langle T_F f,T_F J_\pm f'\rangle}=h_{T_F f}+h_{T_F J_\pm f'}=2h_{T_F f}-\frac{1}{8}(k+2\pm 4m)+\frac{1}{2}\,\delta_{f',T_F f}\,.
\end{equation}
Hence:
\begin{equation}
Q_{(T_F,\psi)}\big(\langle f,J_\pm f'\rangle\big)=h_{(T_F,\psi)}+h_{\langle f,J_\pm f'\rangle}- h_{\langle T_F f,T_F J_\pm f'\rangle}=0\,,
\end{equation}
i.e. these fields are kept in the extension and  organize themselves into orbits. Still, some fields seem not to appear among the off-diagonal field that we would expect. The solutions to this problem is provided by the extension: \textit{fields are pairwise identified}. In fact, as a consequence of (\ref{temp fp of excep curr}), two fields related by the action of (\ref{formual for simple currents in orb}) are mapped into each other by $(T_F,\psi)$ and hence are identified by the currents (\ref{formual for exceptional simple currents}) in the extension.

What happens in determining the fixed points of the exceptional currents is the following. Start with a field $f$ which has $l$-label equal to $\frac{k}{2}$ and apply $J_\pm$ on $f$, recalling that $J^4_\pm=1$ and $J^2_\pm=T_F$ for NS-type currents,
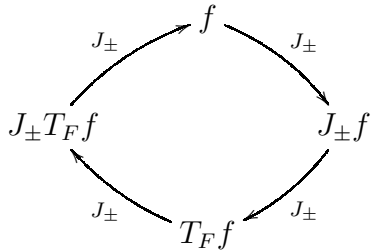
\begin{displaymath}
\xymatrix{
& f \ar@/^/[dr]^{J_\pm} & \\
J_\pm T_F f \ar@/^/[ur]^{J_\pm} & & J_\pm f \ar@/^/[dl]^{J_\pm} \\
& T_F f \ar@/^/[ul]^{J_\pm} &
}
\end{displaymath}
as shown in the graph. The four fields organize themselves pairwise into two $J_\pm$-orbits which are related by the action of $T_F$, or better of $(T_F,\psi)$. In fact, from the fusion rules of $(T_F,\psi)$ with off-diagonal fields it follows that
\begin{equation}
(T_F,\psi)\cdot\langle f,J_\pm f\rangle=\langle T_F f, J_\pm T_F f\rangle\,.
\end{equation}
Each $J_\pm$-orbit has the same form as (\ref{offdiag f.p. of except curr}). 
In the $(T_F,\psi)$-extension they are identified and becomes fixed points of the exceptional simple currents (\ref{formual for exceptional simple currents}). \\
Similarly, we can organize the fields differently. For instance, by starting from the $J_\pm$-orbit $\langle f,J_\pm T_F f\rangle$, we have
\begin{equation}
(T_F,\psi)\cdot\langle f,J_\pm T_F f\rangle=\langle T_F f, J_\pm f\rangle\,,
\end{equation}
where we used $T_F^2=1$. The same argument holds if we start from any $J_\pm$-orbit of two consecutive fields in the graph above: the $(T_F,\psi)$-extension will always identify it with the remaining orbit.

In the next subsection we give and explicit example corresponding to the ``easy'' case of minimal models at level two.

\subsection{$k=2$ Example}
In order to better visualize the structure of exceptional simple currents and their fixed points, let us consider the $k=2$ case, where we permute two $N=2$ minimal models at level two. This case is easy enough to be worked out explicitly, but complicated enough to show all the desired properties. This minimal model has 24 fields (12 in the R sector and 12 in the NS sector), of which 16 simple currents. Using \cite{Klemm:1990df,Borisov:1997nc,Maio:2009tg}, its permutation orbifold has got 372 fields, of which 32 simple currents coming from diagonal (symmetric and anti-symmetric) combinations of the original simple currents. The ones with (half-)integer spin have generically got fixed points which we know how to resolve \cite{Maio:2009tg}.

In the $(T_F,\psi)$-extended orbifold theory, the exceptional currents with fixed points are
\begin{equation}
\langle J_\pm,T_F\cdot J_\pm\rangle_\alpha\,\,,\qquad \alpha=0,\,1\,,
\end{equation}
with
\begin{equation}
J_+=(0,2,0)\qquad{\rm and}\qquad J_-=(0,-2,0)\,.
\end{equation}
Their off-diagonal fixed points are of the form
\begin{equation}
\langle f,J_\pm f'\rangle\,,
\end{equation}
with $f$ and $J_\pm f'$ given by
\begin{eqnarray}
f=(1,1,0) &{\rm and}& f'=(1,-1,0)\nonumber\\
f=(1,2,1) &{\rm and}& f'=(1,0,1)\nonumber\\
f=(1,-1,0) &{\rm and}& f'=(1,1,2)\nonumber\\
f=(1,2,1) &{\rm and}& f'=(1,0,-1)\nonumber
\end{eqnarray}
To these, we still have to add the twisted fixed points, but we know already exactly what they are. 
One can observe that some fields appear twice, e.g. $(1,2,1)$, and other fields never appear, e.g. $(1,2,-1)$. This can be easily explained.
The reason why some of them appear more than once is because $f$ and $f'$ can have either equal or different $s$-values ($J_\pm$ only acts on the $m$-values).\\
Similarly, some fields are identified by the $(T_F,\psi)$-extension and hence they \textit{seem} never to appear. For example, the off-diagonal field $\langle(1,2,-1),(1,0,1)\rangle$ seems not to be there, but it is actually identified  with $\langle(1,2,1),(1,0,-1)\rangle$, which appears in the last line of the list above; similarly $\langle(1,2,-1),(1,0,-1)\rangle$ seems again not to be there as well, but it is identified with $\langle(1,2,1),(1,0,1)\rangle$ which is there in the second line of the same list.

More in general, this is a consequence of (\ref{temp fp of excep curr}). In the present situation we see this explicitly. Let us look at the  current 
\begin{equation}
\langle(0,2,0),(0,2,2)\rangle
\end{equation}
in the permutation orbifold and compute its fusion rules with the off-diagonal field $\langle(1,2,-1),(1,0,1)\rangle$:
\begin{equation}
\label{example fusion}
\langle(0,2,0),(0,2,2)\rangle\cdot \langle(1,2,-1),(1,0,1)\rangle= 
\langle(1,2,-1),(1,0,1)\rangle+\langle(1,2,1),(1,0,-1)\rangle\,.
\end{equation}
We see the appearance of the second term on the r.h.s., which is also an off-diagonal field, so we are led to ask about its fusion as well:
\begin{equation}
\label{example fusion 2}
\langle(0,2,0),(0,2,2)\rangle\cdot \langle(1,2,1),(1,0,-1)\rangle= 
\langle(1,2,-1),(1,0,1)\rangle+\langle(1,2,1),(1,0,-1)\rangle\,,
\end{equation}
which is exactly the same as the first one. However, observe that the current $(T_F,\psi)$ relates the two terms on both r.h.s.'s:
\begin{eqnarray}
(T_F,\psi)\cdot \langle(1,2,-1),(1,0,1)\rangle&=&\langle(1,2,1),(1,0,-1)\rangle\nonumber\\
(T_F,\psi)\cdot \langle(1,2,1),(1,0,-1)\rangle&=&\langle(1,2,-1),(1,0,1)\rangle\,.
\end{eqnarray}
Then, they form one orbit in the $(T_F,\psi)$-extension and, since they have integer monodromy charge, this off-diagonal orbit survives the projection. Due to (\ref{example fusion}) and (\ref{example fusion 2}), this orbit becomes an off-diagonal fixed point of the exceptional current.

As a comment, we remark that it is not known at the moment how to resolve these fixed points. The reason is that they are fixed points of an off-diagonal current for which there is no solution yet, unlike for the fixed points of diagonal currents for which the solution exists and was provided in \cite{Maio:2009tg}.

\section{Orbit structure for $N=2$ and $N=1$}

Here we want to summarize the simple current orbits for theories considered here, and give the analogous results
for $N=1$ minimal models for comparison. Most of the construction, and in particular the definition of the six kinds
of CFT listed in the introduction works completely analogously for $N=2$ and $N=1$.
The worldsheet supercurrent, originating from the diagonal field $\langle 0,T_F \rangle$, comes in both cases from a 
fixed point. However, a novel feature occurring for $N=1$ but not for $N=2$ is that this supercurrent itself has  fixed points 
whose resolution requires additional data. 

Another important difference between the $N=2$ and $N=1$ permutation orbifolds is that in the latter case the supersymmetric and
the non-supersymmetric orbifold (the extensions of the BHS orbifold by $(T_F,1)$ or $(T_F,0)$ respectively) have a different
number of primaries, whereas for $N=2$ this is the same.

The simple current groups of all these theories are as described below. A few currents always play a special r\^ole, namely
\begin{itemize}
\item{The ``un-orbifold" current. This is the current that undoes the permutation orbifold. In the BHS orbifold this is
the anti-symmetric diagonal field $(0,1)$, which has spin-1. If the theories are extended by $(T_F,1)$ or $(T_F,0)$ this
field becomes part of a larger module, but is still the ground state of that module.}
\item{The worldsheet supercurrent(s). This has always weight $\frac32$, and can have fixed points only for $N=1$ (and then
it usually does). The supersymmetric permutation orbifolds always have two of them, which originate from the
split fixed points of the off-diagonal field $\langle 0, T_F \rangle$. Note that this multiplicity, two, has nothing to do with
the number of supersymmetries. The latter is given by the dimension of the ground state of the supercurrent module.
The fusion product of the two supercurrents is always the un-orbifold current. These spin-$\frac32$ currents also occur in the non-supersymmetric
theory $X$, except in that case they generate a $\mathbb{Z}_4$ group, whereas in the supersymmetric case the discrete group they generate is
 $\mathbb{Z}_2\times \mathbb{Z}_2$.}
\item{The Ramond ground state simple currents. These exist only for the $N=2$ and not for the $N=1$ superconformal models.}
\end{itemize}

In the following we call a fixed point ``resolvable" if we have explicit formulas for the fixed point resolution matrices,
and unresolvable otherwise. Therefore, ``unresolvable" does not mean that the fixed points cannot be resolved in principle, but
simply that it is not yet known how to do it.
Note that the choices of generators of discrete
groups described below are not unique, but we made convenient choices.  As much as possible, we try to choose the special
currents listed above as generators of the discrete group factors.
\begin{itemize}
\item{$N=2,\,\, k=1\mod 2$. 
\begin{itemize}
\item{The minimal models have a simple current group $\mathbb{Z}_{4k+8}$. As its generator one can take the
Ramond ground state simple current. The power $2k+4$ of this generator is the worldsheet supercurrent. None of the
simple current has fixed points.}
\item{The supersymmetric permutation orbifold has a group structure $\mathbb{Z}_{4k+8} \times \mathbb{Z}_2$. The first factor is generated by
the Ramond ground state simple current. The power $2k+4$ of this generator is the un-orbifold current. This is the only current that has fixed points, which are resolvable. The factor $\mathbb{Z}_2$
is generated by the worldsheet supercurrent.}
\item{The non-supersymmetric permutation orbifold $X$  also has a group structure $\mathbb{Z}_{4k+8} \times \mathbb{Z}_2$. The spin-$\frac32$
fields originating from the diagonal field $\langle 0,T_F \rangle$ have order 4, and generate a $\mathbb{Z}_{k+2}$ subgroup of  $\mathbb{Z}_{4k+8}$.
The order-two element of $\mathbb{Z}_{4k+8}$ is, just as above, the un-orbifold current. Also in this case it has resolvable fixed points. 
}
\end{itemize}
}
\item{$N=2, \,\,k=0\mod 4$. 
\begin{itemize}
\item{The minimal models have a simple current group $\mathbb{Z}_{2k+4}\times \mathbb{Z}_2$. As the generator of the first factor one can take the
Ramond ground state  simple current, and the worldsheet supercurrent can be used as the generator of the second. 
The middle element of the  $\mathbb{Z}_{2k+4}$ factor is an integer spin current with resolvable fixed points.}
\item{The supersymmetric permutation orbifold has a group structure $\mathbb{Z}_{2k+4} \times \mathbb{Z}_2 \times \mathbb{Z}_2$. The first factor is generated by
the Ramond ground state simple current. The second factor by the 
un-orbifold current. The last factor is generated by the worldsheet supercurrent. The middle element of the first factor and
the generator of the second factor, as well as their product have resolvable fixed points.}
\item{The non-supersymmetric permutation orbifold $X$  has a group structure $\mathbb{Z}_{2k+4} \times \mathbb{Z}_4$. The spin-$\frac32$
fields originating from the diagonal field $\langle 0,T_F \rangle$ have order 4 
can be chosen as generators of the $\mathbb{Z}_4$ factor. 
There are three non-trivial currents with resolvable fixed points, which have the same origin (in terms of minimal model fields)
 as the ones in the supersymmetric orbifold.
}
\end{itemize}
}
\item{$N=2,\,\, k=2\mod 4$. 
\begin{itemize}
\item{The minimal models have a simple current group $\mathbb{Z}_{2k+4}\times \mathbb{Z}_2$. The structure is exactly as for $k=0\mod4$.}
\item{The supersymmetric permutation orbifold has a group structure $\mathbb{Z}_{2k+4} \times \mathbb{Z}_2 \times \mathbb{Z}_2$. One can choose
the same generators as above for $k=0\mod4$. The fixed point structure is also identical, except that there are four
additional currents with unresolvable fixed points. These four currents are the two order 4 currents of $\mathbb{Z}_{2k+4}$ multiplied with
each of the two world-sheet supercurrents.} 
\item{The non-supersymmetric permutation orbifold $X$  has a group structure $\mathbb{Z}_{2k+4} \times \mathbb{Z}_4$. As in the
supersymmetric case, there are three non-trivial currents with resolvable fixed points, and four with unresolvable fixed points.  These currents have
the same origin as those of the supersymmetric orbifold.
}
\end{itemize}
}
\item{$N=1,\,\, k=1\mod 2$. 
\begin{itemize}
\item{The minimal models have a simple current group $\mathbb{Z}_{2}$, generated by the worldsheet supercurrent. This current has
resolvable fixed points.}
\item{The supersymmetric permutation orbifold has a group structure $\mathbb{Z}_{2} \times \mathbb{Z}_2$. The two factors can be
generated by the un-orbifold current and by the worldsheet current. The fourth element also has spin-$\frac32$, and is an alternative worldsheet
supercurrent. The un-orbifold current has resolvable fixed points, the supercurrents have unresolvable fixed points. }
\item{The non-supersymmetric permutation orbifold $X$  has a group structure $\mathbb{Z}_{8}$. The order-2 element in this
subgroup is the un-orbifold current, which has resolvable fixed points. None of the other currents have fixed points.}
\end{itemize}
}
\item{$N=1,\,\, k=0\mod 2$. 
\begin{itemize}
\item{The minimal models have a simple current group $\mathbb{Z}_{2}\times \mathbb{Z}_2$. All currents have resolvable fixed points.
One of them is the worldsheet supercurrent.}
\item{The supersymmetric permutation orbifold has a group structure $\mathbb{Z}_{2} \times \mathbb{Z}_2 \times \mathbb{Z}_2$. Two of the three factors
are generated by the un-orbifold current and one of the worldsheet supercurrents. All currents have fixed points, and for four of them,
including the supersymmetry generators, they are unresolvable.}
\item{The non-supersymmetric permutation orbifold $X$  has a group structure $\mathbb{Z}_{4} \times \mathbb{Z}_2$. All currents have
fixed points, and for four of them they are unresolvable. 
}
\end{itemize}
}
\end{itemize}

\section{Conclusion}
In this paper we study permutation and extensions of $N=2$ minimal models at arbitrary level $k$. These models are very interesting for several reason: not only because they are non-trivial solvable conformal field theories, but also because they are the building blocks of Gepner models which have some relevance in string theory phenomenology.

Our main points are two. First of all, a new structure arises relating conformal field theories built out of minimal models. Starting from the tensor product we perform $\mathbb{Z}_2$-orbifold and extension in both possible orders, generating in this way new CFT. Some of them are easily recognizable, such as the $N=2$ supersymmetric orbifold obtained by extending the standard permutation orbifold by the current $(T_F,1)$. Some others are however not known, like the CFT that we have denoted by $X$, obtained by extending the orbifold by $(T_F,0)$. 

Secondly, unexpected off-diagonal simple currents appear due to the interplay of the orbifold and the extension procedure. Sometimes they have fixed points that need to be resolved. However, because they are related to off-diagonal currents, we do not know how to resolve them at the moment.

The most natural and immediate application of our method is to consider permutation orbifolds in Gepner models, which are built out of $N=2$ supersymmetric minimal models. 
The results of this paper allow us to consider permutation orbifold building blocks in combination with minimal models to
build new closed string theories. These closed string theories can be the starting point of orientifold model building as well
as heterotic model building. 
Since Gepner models allow a geometric interpretation as heterotic string theory compactifications on a Calabi-Yau manifold \cite{Candelas:1985en}, our results will extend the work of \cite{Fuchs:1991vu} regarding permutations in Gepner-type superstrings. But we can go
a lot further than that, because we can consider subgroups of the canonical $SO(10)$ gauge group, break the remnants of space-time
and  world-sheet supersymmetry on the bosonic side, and combine all this with heterotic weight lifting and B-L lifting, following \cite{Schellekens:1989wx,GatoRivera:2010gv,GatoRivera:2010xn,GatoRivera:2010xna}. This application also  provides important tests
on the structure of the CFT building blocks. Anomaly cancellation is a very unforgiving constraint in these more general heterotic strings
(as opposed to (2,2) models with families of $(27)$'s of $E_6$, where it is automatic). We have already performed the first successful tests of
the results of the present paper applied to heterotic strings, but we will present the results elsewhere \cite{Maio:5}. 

The application to orientifold model building is also possible, and it will be interesting to see if this extends the set of
realizable brane configurations, and/or enhances the possibilities for tadpole cancellation. However, we still face one limitation here, since
we cannot use permutation orbifolds with $k\not=2 \mod 4$. Perhaps this can be evaded by simply not using the simple currents
that have unresolvable fixed points. However, the unresolved fixed points occur for currents that are products of powers of the Ramond ground state
simple current and the worldsheet supercurrent. Both of these components are certainly needed separately, and it is not immediately obvious
if there are any MIPFs where the unresolved fixed points can be avoided. Even if $k\not=2 \mod 4$ cannot be used,
 this still leaves us with three-quarters of the $N=2$ minimal permutation orbifolds, including permutations
of the factors of the ``quintic" $(3,3,3,3,3)$. We hope to report on this application of our results in the future.

\section*{Acknowledgments}
This research is supported by the Dutch Foundation for Fundamental Research of Matter (FOM)
as part of the program STQG (String Theory and Quantum Gravity, FP 57). This work has been partially 
supported by funding of the Spanish Ministerio de Ciencia e Innovaci\'on, Research Project
FPA2008-02968, and by the Project CONSOLIDER-INGENIO 2010, Programme CPAN (CSD2007-00042). \\
M.M. would like to thank Thomas Quella for discussions about various topics, 
some of them (but not all) also related to this paper, 
as well as Ole Warnaar for email exchange about the string functions.

\appendix

\section{Twisted-fields orbits of the $(0,1)$-current}
In this appendix we want to prove that in any permutation orbifold the simple current $(0,1)$ (anti-symmetric representation of the identity) always couples a twisted field to its own (un)excited partner, i.e.
\begin{equation}
\widehat{(p,0)} \stackrel{(0,1)}{\leftrightarrow} \widehat{(p,1)}\,.
\end{equation}
To prove this, let use compute the fusion coefficients:
\begin{equation}
(0,1)\cdot \widehat{(p,\xi)} = \sum_K N_{(0,1)\widehat{(p,\xi)}}^{\phantom{(0,1)\widehat{(p,\xi)}}K} (K)\,,
\end{equation}
where the sum runs aver all the fields $K$ in the orbifold. By Verlinde's formula \cite{Verlinde:1988sn}:
\begin{eqnarray}
N_{(0,1)\widehat{(p,\xi)}}^{\phantom{(0,1)\widehat{(p,\xi)}}K} 
&=&  \sum_{N} 
\frac{S_{(0,1)N}S_{\widehat{(p,\xi)}N}S_{\phantom{\dagger}N}^{\dagger \phantom{N}K}}{S_{(0,0)N}}=
\nonumber\\  &=&  
\sum_{\langle i,j\rangle}
\frac{S_{(0,1)\langle i,j \rangle}S_{\widehat{(p,\xi)}\langle i,j \rangle}S_{\phantom{\dagger}\langle i,j \rangle}^{\dagger\phantom{\langle i,j \rangle}K}}{S_{(0,0)\langle i,j \rangle}}+
\nonumber\\ &+&  
\sum_{(j,\chi)}
\frac{S_{(0,1)(j,\chi)}S_{\widehat{(p,\xi)}(j,\chi)}S_{\phantom{\dagger}(j,\chi)}^{\dagger\phantom{(j,\chi)}K}}{S_{(0,0)(j,\chi)}}+
\nonumber\\ &+& 
\sum_{\widehat{(j,\xi)}}
\frac{S_{(0,1)\widehat{(j,\chi)}}S_{\widehat{(p,\xi)}\widehat{(j,\chi)}}S_{\phantom{\dagger}\widehat{(j,\chi)}}^{\dagger\phantom{\widehat{(j,\chi)}}K}}{S_{(0,0)\widehat{(j,\chi)}}}
\nonumber\,.
\end{eqnarray}
Now use the orbifold $S$ matrix (\ref{BHS}): the first line automatically vanishes, since the $S^{BHS}$ vanishes when one entry is a twisted field and the other one is off-diagonal. The other two lines give
\begin{equation}
N_{(0,1)\widehat{(p,\xi)}}^{\phantom{(0,1)\widehat{(p,\xi)}}K} =
\frac{1}{2}\sum_{\chi=0}^1\sum_j e^{i\pi\chi}\, S_{pj} \cdot S_{(j,\chi)}^{\star\phantom{(j,\chi)}K} -
\frac{1}{2}\sum_{\chi=0}^1\sum_j e^{i\pi(\xi+\chi)}\, P_{pj} \cdot S_{\widehat{(j,\chi)}}^{\star\phantom{\widehat{(j,\chi)}}K}\,.
\nonumber
\end{equation}
The two contributions both vanish if $K$ is of diagonal type or of off-diagonal type, as one can easily verify by using (\ref{BHS}). On the other hand, if $K$ is of twisted type, we find a non-vanishing answer that can be written as
\begin{equation}
N_{(0,1)\widehat{(p,\xi)}}^{\phantom{(0,1)\widehat{(p,\xi)}}\widehat{(k,\eta)}} =
\frac{1}{2}\, \delta_p^k\,(1-e^{i\pi(\xi-\eta)})=\delta_p^k\,\delta_{\xi+1}^\eta\,.
\end{equation}
Here we have used unitarity of the $S$ and $P$ matrices. In other words,
\begin{equation}
(0,1)\cdot \widehat{(p,0)} = \widehat{(p,1)}\,,
\end{equation}
as well as the other way around, being the current $(0,1)$ of order two.

\section{Fusion rules of $\langle 0,T_F\rangle$ and corresponding split fields}
In this section we would like to show that the fusion coefficients of $\langle 0,T_F\rangle$ with itself, before and after the $(T_F,\psi)$-extension, do not depend on the sign choice for the coefficients $A$ and $C$ appearing in the $S^J$ ansatz (\ref{MS ansatz}). In particular, the intrinsic ambiguity related to the freedom of ordering twisted fields (i.e. which one we label by $\chi=0$ and which one by $\chi=1$) should not make any difference in the calculation of the fusion rules. 
The calculation is straightforward and relatively short before making the $(T_F,\psi)$-extension, since it involves only the BHS $S$ matrix: we will describe it in detail. 

However, after taking the $(T_F,\psi)$-extension, the full extended $S$ matrix must be used. This means that the BHS $S$ matrix appears together with the $S^{(T_F,\psi)}$ matrix; moreover, fixed point resolution implies that the fixed points of $(T_F,\psi)$ are split, hence there will be twice their number, while non-fixed points form orbits and only half of them will be independent.
The calculation in this case is lengthy and more involved, so we will only point out where the sign ambiguities mentioned above could (but will not) play a role.

\subsection{Before $(T_F,\psi)$-extension}
The quantity that we want to compute is
\begin{equation}
\langle 0,T_F\rangle \cdot \langle 0,T_F\rangle =
\sum_{K} N_{\langle 0,T_F\rangle\langle 0,T_F\rangle}^{\phantom{\langle 0,T_F\rangle\langle 0,T_F\rangle}K} (K)\,,
\end{equation}
where the sum runs over all the fields $K$ of the permutation orbifold. The quantity $N_{\langle 0,T_F\rangle\langle 0,T_F\rangle}^{\phantom{\langle 0,T_F\rangle\langle 0,T_F\rangle}K}$ is given by Verlinde's formula \cite{Verlinde:1988sn}
\begin{equation}
N_{\langle 0,T_F\rangle\langle 0,T_F\rangle}^{\phantom{\langle 0,T_F\rangle\langle 0,T_F\rangle}K}=
\sum_{N} \frac{S_{\langle 0,T_F\rangle N}S_{\langle 0,T_F\rangle N}S_{\phantom{\dagger}N}^{\dagger \phantom{N}K}}{S_{(0,0)N}}\,.
\end{equation}
Let us start with the case that $K$ is a diagonal field, $K=(k,\chi)$, and use the BHS expression for the orbifold $S$ matrix:
\begin{eqnarray}
N_{\langle 0,T_F\rangle\langle 0,T_F\rangle}^{\phantom{\langle 0,T_F\rangle\langle 0,T_F\rangle}(k,\chi)}&=&
\sum_{m<n}\frac{(S_{0m}S_{T_F, n}+S_{0n}S_{T_F, m})^2\cdot(S^\star_{mk}S^\star_{nk})}{S_{0m}S_{0n}}+\nonumber\\
&+&
\sum_{\phi=0}^1\sum_i\frac{(S_{0i}S_{T_F, i})^2\cdot(\frac{1}{2}S^{\star 2}_{ik})}{(\frac{1}{2}S^2_{0i})}+
0\,.\nonumber
\end{eqnarray}
The zero in the second line comes from the twisted contribution, since from the BHS formula $S_{\langle mn\rangle \widehat{(i,\chi)}}=0$. The sum over $\phi$ gives a factor of $2$ in the diagonal contribution. In the first sum we can use
\begin{equation}
\sum_{m,\,n}=2\sum_{m<n}+\sum_{m=n}\,.
\end{equation}
The sum $\sum_{m=n}$ will cancel the diagonal contribution. Eventually we are left only with three terms coming from expanding the square in the sum over $m$ and $n$. The two sums are now independent and factorize:
\begin{eqnarray}
N_{\langle 0,T_F\rangle\langle 0,T_F\rangle}^{\phantom{\langle 0,T_F\rangle\langle 0,T_F\rangle}(k,\chi)}&=&
\frac{1}{2}\sum_m S^\star_{mk}S_{0m}\sum_n \frac{S^\star_{nk}S^2_{T_F, n}}{S_{0n}}+
\frac{1}{2}\sum_n S^\star_{nk}S_{0n}\sum_m \frac{S^\star_{mk}S^2_{T_F, m}}{S_{0m}}+\nonumber\\
&+&
\sum_m S^\star_{mk}S_{0m}\sum_n S^\star_{nk}S_{0n}=\nonumber\\
&=&
\delta_{k,0}N_{T_F T_F}^{\phantom{T_F T_F}k} +\delta_{k,T_F}=\nonumber\\
&=&
\delta_{k,0} +\delta_{k,T_F}\,,
\end{eqnarray}
where we have used the fact that $T_F$ has order two, i.e. $N_{T_F T_F}^{\phantom{T_F T_F}k}=\delta_{k,0}$. Note that the answer does not depend on $\chi$.

We can now repeat the same steps in the case that $K$ is off-diagonal, $K=\langle k_1,k_2\rangle$ (with $k_1<k_2$). 
We get:
\begin{equation}
N_{\langle 0,T_F\rangle\langle 0,T_F\rangle}^{\phantom{\langle 0,T_F\rangle\langle 0,T_F\rangle}\langle k_1,k_2\rangle}\propto \delta_{0,k_1}\cdot\delta_{0,k_2}=0\,,
\end{equation}
since $k_1\neq k_2$.

Similarly, for $K$ twisted, $K=\widehat{(k,\chi)}$:
\begin{equation}
N_{\langle 0,T_F\rangle\langle 0,T_F\rangle}^{\phantom{\langle 0,T_F\rangle\langle 0,T_F\rangle}\widehat{(k,\chi)}}=0+0+0=0\,,
\end{equation}
where the first and third contributions vanish because $S_{\langle mn\rangle\widehat{(i,\chi)}}=0$ in the BHS $S$ matrix, while the second one vanishes because $\sum_{\phi=0}^1 e^{i\pi\phi}=0$.

Putting everything together we have the following fusion rules for $\langle 0,T_F\rangle$ with itself before the $(T_F,\psi)$-extension:
\begin{equation}
\langle 0,T_F\rangle \cdot \langle 0,T_F\rangle=
(0,0)+(0,1)+(T_F,0)+(T_F,1)\,.
\end{equation}

\subsection{After $(T_F,\psi)$-extension}
After the extension by $(T_F,\psi)$, the off-diagonal field $\langle 0,T_F\rangle$ becomes a simple current. Moreover, since it is fixed by $(T_F,\psi)$, as well as $(0,\psi)$, it gets split and originates two simple currents, $\langle 0,T_F\rangle_\alpha$ with $\alpha=0,\,1$.

In order to compute the fusion rules between $\langle 0,T_F\rangle_\alpha$ and $\langle 0,T_F\rangle_\beta$, we need to know the full $S$ matrix of the extension. It is given by \cite{Fuchs:1996dd}
\begin{equation}
\tilde{S}_{a_\alpha b_\beta}= {\rm Const} \cdot (S_{ab}+(-1)^{\alpha+\beta}S^{(T_F,\psi)}_{ab})\,.
\end{equation}
Here, $S_{ab}$ is the BHS $S$ matrix and $S^{(T_F,\psi)}_{ab}$ is the fixed-point resolution matrix $S^J$ corresponding to the current $J=(T_F,\psi)$. The overall constant is a group-theoretical factor such that
\begin{equation}
{\rm Const}=
\left\{
\begin{array}{lc}
\frac{1}{2}  & {\rm if\,\,both\,\,}a\,\&\,b\,\,{\rm are\,\,fixed\,\,points}\\
1 & {\rm if\,\,either\,\,}a\,{\rm or}\,b\,\,({\rm not\,\,both})\,\,{\rm is\,\,fixed\,\,point}\\
2 & {\rm if\,\,neither\,\,}a\,\&\,b\,\,{\rm are\,\,fixed\,\,points}
\end{array}
\right.
\end{equation}
As mentioned in the paper, the $S^{(T_F,\psi)}_{ab}$ in the untwisted sector vanishes, because $T_F$ does not have fixed points.

We want to compute:
\begin{equation}
\langle 0,T_F\rangle_\alpha\cdot\langle 0,T_F\rangle_\beta=
\sum_Q N_{\langle 0,T_F\rangle_\alpha \langle 0,T_F\rangle_\beta}^{\phantom{\langle 0,T_F\rangle_\alpha \langle 0,T_F\rangle_\beta}Q} (Q)\,,
\end{equation}
where
\begin{equation}
N_{\langle 0,T_F\rangle_\alpha \langle 0,T_F\rangle_\beta}^{\phantom{\langle 0,T_F\rangle_\alpha \langle 0,T_F\rangle_\beta}Q}=
\sum_{N} \frac{\tilde{S}_{\langle 0,T_F\rangle_\alpha N}\tilde{S}_{\langle 0,T_F\rangle_\beta N}\tilde{S}_{\phantom{\dagger}N}^{\dagger \phantom{N}Q}}{\tilde{S}_{(0,0)N}}\,.
\end{equation}

Consider $Q$ to be diagonal, $Q=(q,\chi)$. Diagonal fields are never fixed points of $(T_F,\psi)$, hence if the $S^{(T_F,\psi)}$ has at least one diagonal entry it vanishes. Then we have: 
\begin{eqnarray}
N_{\langle 0,T_F\rangle_\alpha \langle 0,T_F\rangle_\beta}^{\phantom{\langle 0,T_F\rangle_\alpha \langle 0,T_F\rangle_\beta}(q,\chi)}
&=&
\sum_{N} \frac{\tilde{S}_{\langle 0,T_F\rangle_\alpha N}\tilde{S}_{\langle 0,T_F\rangle_\beta N}\tilde{S}_{\phantom{\dagger}N}^{\dagger \phantom{N}(q,\chi)}}{\tilde{S}_{(0,0)N}}=\\
&=&
\sum_{\langle mn \rangle} \frac{\tilde{S}_{\langle 0,T_F\rangle_\alpha \langle mn \rangle}\tilde{S}_{\langle 0,T_F\rangle_\beta \langle mn \rangle}\tilde{S}_{\phantom{\dagger}\langle mn \rangle}^{\dagger \phantom{\langle mn \rangle}(q,\chi)}}{\tilde{S}_{(0,0)\langle mn \rangle}}+\nonumber\\
&+&
\sum_{(p,\phi)} \frac{\tilde{S}_{\langle 0,T_F\rangle_\alpha (p,\phi)}\tilde{S}_{\langle 0,T_F\rangle_\beta (p,\phi)}\tilde{S}_{\phantom{\dagger}(p,\phi)}^{\dagger \phantom{(p,\phi)}(q,\chi)}}{\tilde{S}_{(0,0)(p,\phi)}}+\nonumber\\
&+&
\sum_{\gamma=0}^1\sum_{\widehat{(p,\phi)}_\gamma} \frac{\tilde{S}_{\langle 0,T_F\rangle_\alpha \widehat{(p,\phi)}_\gamma}\tilde{S}_{\langle 0,T_F\rangle_\beta \widehat{(p,\phi)}_\gamma}\tilde{S}_{\phantom{\dagger}\widehat{(p,\phi)}_\gamma}^{\dagger \phantom{\widehat{(p,\phi)}_\gamma}(q,\chi)}}{\tilde{S}_{(0,0)\widehat{(p,\phi)}_\gamma}}\,.\nonumber
\end{eqnarray}
Let us stress a few points here. First, the sum over $\langle mn \rangle$ is symbolic: we must consider both the situations when $\langle mn \rangle$ is a fixed point of $(T_F,\psi)$ (in which case it will carry an extra label $\langle mn \rangle_\gamma$, with $\gamma=0$ or $1$) and when it is just an orbit representative (in which case we should not include its partner $\langle T_F m,T_F n \rangle$ in the sum in order to avoid double counting). \\
Diagonal fields are always orbit representatives, while twisted fields are always fixed points. In principle, the $S^{(T_F,\psi)}$ matrix can appear in the sums over $\langle mn \rangle$ and over $\widehat{(p,\phi)}$, but in practice it only appear in the latter, since it vanishes for untwisted-untwisted entries. So the possible ambiguity might play a role only in the last line. Hence let us have a closer look there. For off-diagonal-twisted entries, the BHS $S$ matrix is identically zero, so we can replace $\tilde{S}$ with $S^{(T_F,\psi)}$, up to the overall constant. Using the ansatz (\ref{MS ansatz}), the contribution to the fusion rules from the last line is then
\begin{eqnarray}
&& 2
\sum_{\begin{array}{c}
\widehat{(p,\phi)}\\
\widehat{(p,\phi)}\,\,{\rm f.p.\,of\,\,}S^{(T_F,\psi)}
\end{array}}
\frac{
\frac{1}{2}(-1)^{\alpha+\gamma}A S_{0p} \cdot \frac{1}{2}(-1)^{\beta+\gamma}A S_{0p} \cdot C^\star\frac{1}{2}e^{-\i\pi\chi} S^{\star}_{pq}
}{
C\frac{1}{2}S_{0p}
}=\nonumber \\
&=&
\frac{1}{2}\,A^2\,\frac{C^\star}{C}\,(-1)^{\alpha+\beta}\,e^{-i\pi\chi}\,\,\cdot
\sum_{\begin{array}{c}
\widehat{(p,\phi)}\\
\widehat{(p,\phi)}\,\,{\rm f.p.\,of\,\,}S^{(T_F,\psi)}
\end{array}}
S_{0p}S^\star_{pq}\nonumber\,.
\end{eqnarray}
The sum over $\widehat{(p,\phi)}$ fixed points of $S^{(T_F,\psi)}$ can be computed using Corollary 1 in the Appendix of \cite{Maio:2009tg}. It contains the $\psi$ dependence. What is relevant for our discussion here is the prefactor: there is no ambiguity related to different choices for the coefficients $A$ and $C$, since changing $A\rightarrow -A$ and/or $C\rightarrow -C$ would not alter the result.

The full and exact calculation of the fusion rules  after the $S^{(T_F,\psi)}$-extension is too lengthy to be repeated and we will not do it here. In particular, the cases when $Q$ is off-diagonal or twisted are not very relevant, since then the fusion coefficients vanish identically, as one can check numerically. We simply state the outcome of the complete calculation:
\begin{itemize}
\item For the $(T_F,0)$-extension:
\begin{eqnarray}
\langle 0,T_F\rangle_\alpha \cdot \langle 0,T_F\rangle_\alpha &=& (0,1) \qquad \alpha=0,\,1\nonumber\\
\langle 0,T_F\rangle_\alpha \cdot \langle 0,T_F\rangle_\beta &=& (0,0) \qquad \alpha\neq\beta\,;
\end{eqnarray}
hence $\langle 0,T_F\rangle_\alpha$ is of order four, being $(0,1)\cdot(0,1)=(0,0)$, so it cannot be a supersymmetry current.
\item For the $(T_F,1)$-extension:
\begin{eqnarray}
\langle 0,T_F\rangle_\alpha \cdot \langle 0,T_F\rangle_\alpha &=& (0,0) \qquad \alpha=0,\,1\nonumber\\
\langle 0,T_F\rangle_\alpha \cdot \langle 0,T_F\rangle_\beta &=& (0,1) \qquad \alpha\neq\beta\,;
\end{eqnarray}
hence $\langle 0,T_F\rangle_\alpha$ is of order two, as a supersymmetry current should be.
\end{itemize}
Note that in both cases only a particular diagonal field contributes to the fusion rules, namely the identity, as one could have expected because of the order two of $T_F$.


\end{document}